\documentclass[aps,prb,twocolumn,superscriptaddress]{revtex4-2}
\usepackage{graphicx}
\usepackage{amssymb}
\usepackage{amsmath}
\usepackage{color}

\def\kHgXd{$\kappa$-(BEDT\--TTF)$_4$\-Hg$_{3-\delta}X_8$}

\def\HgBr{\mbox{$\kappa$-Hg$_{3-\delta}$Br$_8$}}
\def\HgCl{\mbox{$\kappa$-Hg$_{3-\delta}$Cl$_8$}}
\def\HgXd{\mbox{$\kappa$-Hg$_{3-\delta}X_8$}}
\def\kCuNCS{$\kappa$-(BEDT\--TTF)$_2$\-Cu(NCS)$_2$}
\def\kCuBr{$\kappa$-(BEDT\--TTF)$_2$\-Cu[N(CN)$_2$]Br}
\def\CuNCS{\mbox{$\kappa$-CuNCS}}
\def\CuBr{\mbox{$\kappa$-CuBr}}
\def\cm{cm$^{-1}$}

\begin{document}

\title{Strange-metal behavior in correlated molecular conductors}

\author{Savita Priya}
\affiliation{1.~Physikalisches Institut, Universit\"{a}t Stuttgart, Pfaffenwaldring 57, 70569 Stuttgart, Germany}

\author{Maxim Wenzel}
\affiliation{1.~Physikalisches Institut, Universit\"{a}t Stuttgart, Pfaffenwaldring 57, 70569 Stuttgart, Germany}

\author{Christian Prange}
\affiliation{1.~Physikalisches Institut, Universit\"{a}t Stuttgart, Pfaffenwaldring 57, 70569 Stuttgart, Germany}

\author{Bertold Rasche}
\affiliation{Institut f{\"u}r Anorganische Chemie, Universit\"{a}t Stuttgart, Pfaffenwaldring 55, 70569 Stuttgart, Germany}

\author{Christian Sindlinger}
\affiliation{Institut f{\"u}r Anorganische Chemie, Universit\"{a}t Stuttgart, Pfaffenwaldring 55, 70569 Stuttgart, Germany}

\author{\\Takuya Kobayashi}
\affiliation{Graduate School of Science and Engineering, Saitama University, Saitama 338-8570, Japan}

\author{Hiromi Taniguchi}
\affiliation{Graduate School of Science and Engineering, Saitama University, Saitama 338-8570, Japan}

\author{Dita Puspita Sari}
\affiliation{Nuclear Structure Research Group, RIKEN Nishina Center, 
Saitama, 351-0198, Japan}

\affiliation{Shibaura Institute of Technology, 
Saitama 337-8570, Japan}

\author{Yasuyuki Ishii}
\affiliation{Shibaura Institute of Technology, 
Saitama 337-8570, Japan}

\author{\\Kazuya Miyagawa}
\affiliation{Department of Applied Physics, University of Tokyo, Tokyo 113-8656, Japan}

\author{Kazushi Kanoda}
\affiliation{1.~Physikalisches Institut, Universit\"{a}t Stuttgart, Pfaffenwaldring 57, 70569 Stuttgart, Germany}
\affiliation{Max Planck Institute for Solid State Research, Heisenbergstra\ss{}e 1, 70569 Stuttgart, Germany}
\affiliation{Department of Advanced Materials Science, University of Tokyo, Tokyo 113-8656, Japan}

\author{Simone Fratini}
\affiliation{Institut N\'{e}el - CNRS and Universit\'{e} Grenoble Alpes, 38042 Grenoble Cedex 9, France}

\author{Martin Dressel}
\affiliation{1.~Physikalisches Institut, Universit\"{a}t Stuttgart, Pfaffenwaldring 57, 70569 Stuttgart, Germany}

\date{\today}

\begin{abstract}
The  $\kappa$-BEDT-TTF salts are prototypical Mott systems, realizing a correlation-driven
electronic transition from a Fer\-mi-liquid state to a Mott insulator.
The metallic compounds with spatially modulated Hg-containing anions,
{$\kappa$-(BEDT-TTF)$_4$\-Hg$_{3-\delta}X_8$} ($X$ = Br, and Cl), however, fall outside this classification.
Our temperature-dependent transport and broad-range optical investigations reveal a  $\rho(T)\propto T$ dependence of the resistivity up to elevated temperatures and a power-law behavior of the frequency-de\-pen\-dent conductivity $\vert\hat{\sigma} (\omega)\vert \propto \omega^{-0.7}$; well-known hallmarks of ``strange metals''.
The linear-in-frequency scattering rate of these strange metals scales to the Planckian dissipation.
We demonstrate that this response only occurs in systems with incommensurate anion layers and establish a direct link to large-scale spatial modulations.
Our conclusions can be extended to other strongly correlated strange-metal compounds.
\end{abstract}

\maketitle

\textit{Introduction---}The label ``strange metal'' marks the breakdown of Fer\-mi-liquid quasiparticle transport:
it is assigned to compounds whose charge transport violates the temperature-dependent power laws of conventional metals and records linear resistivity $\rho(T)\propto T$~\cite{Phillips22,Savitsky25,Zaanen04}. Even though this behavior is frequently observed in doped systems with correlated electrons,
its origin remains one of the most puzzling questions in condensed matter physics.
Theoretical studies have linked it to various effects such as electronic correlations, disorder, magnetic/spin fluctuations, quantum criticality and entanglement, without reaching consensus; but they do agree on the observation of Planckian dissipation~\cite{Phillips22,Patel19,Gleis25,Fratini26,Romero26,Hartnoll22, Michon23,Terletska11,Seibold21,Mazza26,Fratini26np}. It has been suggested that the presence of electronic inhomogeneities could be a decisive factor in the emergence of strange-metal behavior in doped transition-metal superconductors~\cite{Patel23,*Patel25,Gleis25}. However, since a certain degree of structural disorder is unavoidable in these materials, it is not possible to disentangle the effects of correlations, dynamical fluctuations (charge or magnetic or related to superconductivity) or structural disorder from one another.
Thus, the question of the origin of ``strange-metal'' response remains unresolved.

In this regard, low-dimensional molecular compounds, including $\kappa$-BEDT-TTF salts [BEDT-TTF stands for bis\-(ethylene\-di\-thio)-tetra\-thia\-fulvalene] offer ideal model systems for investigating complex interactions because they possess high crystal quality, a simple Fermi surface,
extended and precise variability for tuning competing interactions and frustration by chemical or physical means~\cite{ToyotaBook,Powell11,Dressel20,Riedl22}.
Above all, the absence of chemical doping ensures that the electronic system is not affected by extrinsic disorder.
Though defects can be intentionally introduced through irradiation~\cite{Sasaki12c,Yamamoto20}.

\begin{figure*}[t]
    \centering
        \includegraphics[width=0.85\linewidth]{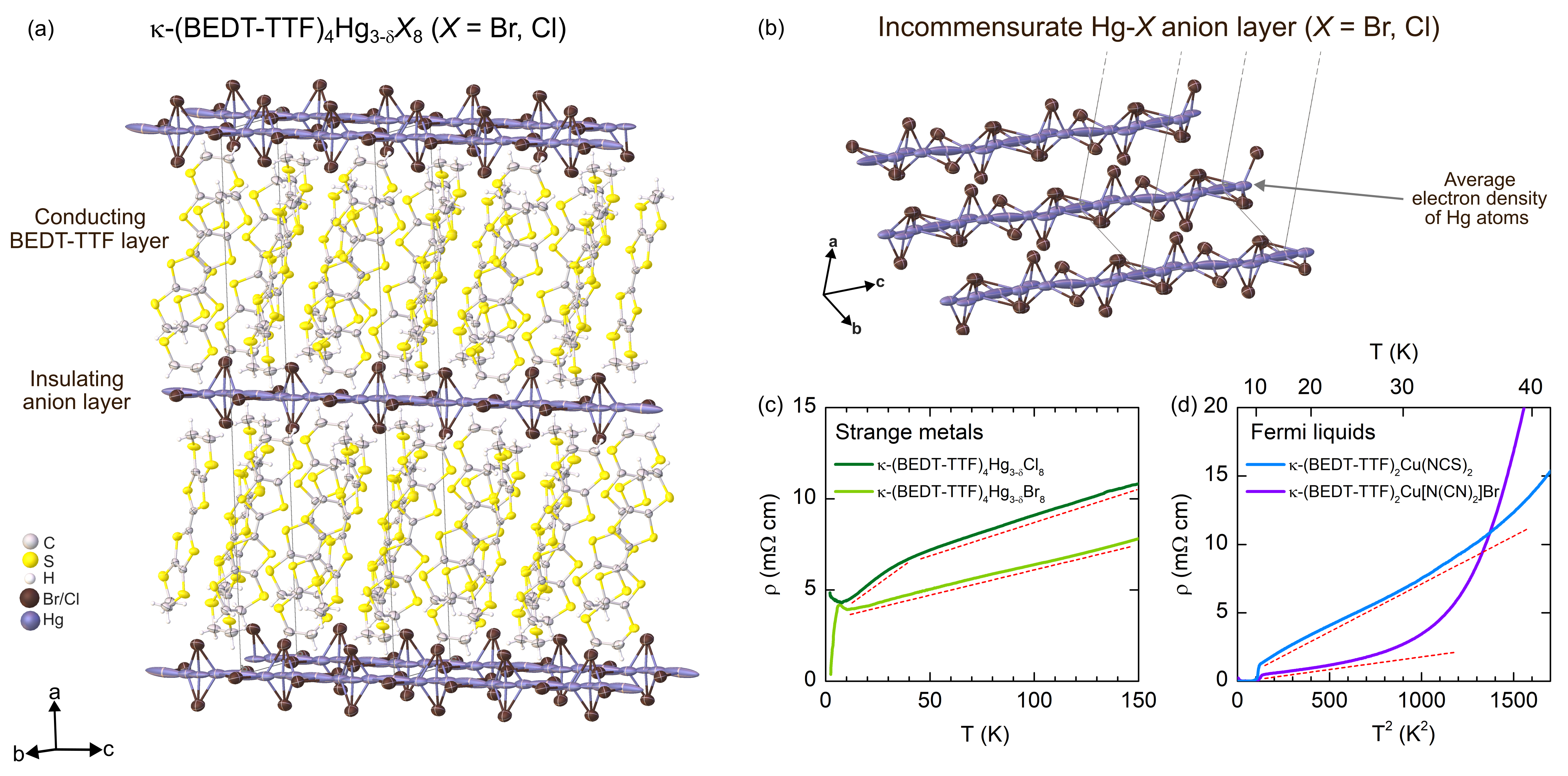}
        \caption{(a)~Crystal structure of $\kappa$-(BEDT-TTF)$_4$Hg$_{3-\delta}X_8$, composed of conducting BEDT-TTF layers separated by insulating anion layers. (b)~Incommensurate Hg-$X$ anion layer in \HgXd{} viewed from a different angle. (c)~Temperature-dependent dc resistivity of the strange metals, demonstrating $\rho(T)\propto T$; and (d)~electrical resistivity {\it vs.} $T^2$ plot for two correlated Fermi liquids, revealing the $\rho(T)\propto T^2$ response. The red dashed lines are guides to the eye.}
     \label{fig:structure+rho} 
\end{figure*}

We study the electrodynamics of two charge transfer salts,
\kHgXd~\cite{Lyubovskaya91,Lyubovskii96} ($X$ = Br, and Cl, abbreviated \HgBr{} and \HgCl;
$\delta$ = 0.10 and 0.18, respectively), which are composed of incommensurate anion layers as illustrated in Fig.~\ref{fig:structure+rho}(a,b) \footnote{The estimation of the
degree of imperfection (i.e. amount of vacancies concentration during crystal growth) depends
on the applied model for structural analysis:  $\delta$ = 0.11-0.25. Details are given in the Supplemental Materials \cite{SM}.}.
The conducting BEDT-TTF cation layer is isostructural to the other $\kappa$-BEDT-TTF compounds, however,
the anion structure is well distinct.
Oike and collaborators reported deviations from canonical Fer\-mi-liquid properties in \HgBr\ \cite{Oike17,Suzuki22,Oike24}.

Here we investigate the response of itinerant carriers in their metallic states over a large temperature and spectral range using broadband infrared spectroscopy; see Supplemental Material \cite{SM} for details on experiments and analysis.
Our findings on \kHgXd\ are compared with the low-temperature properties of the Fermi liquids \kCuBr{} and \kCuNCS{} (\CuBr{} and \CuNCS).

\textit{Results---}Fig.~\ref{fig:structure+rho}(c) shows the electrical resistivity $\rho(T)$ of \kHgXd\ measured as a function of temperature. Both systems exhibit metallic behavior with a
linear-in-$T$ increase  up to elevated temperatures and comparable slopes. In the case of \HgBr\ the remarkable linear behavior is only limited by a weak upturn above the superconducting transition at $T_c\approx 3.5$~K. Our transport results for \HgBr\ are consistent with the trends under pressure and magnetic fields~\cite{Suzuki22}. For \HgCl, a similar upturn is observed as $T\to 0$ but no superconductivity, and the slope slightly changes below 40~K.
Most importantly, a $\rho(T)\propto T^2$ dependence, expected for a Fermi liquid, is absent for both \HgXd\ compounds. Finding $\rho(T)\propto T$ in such a wide temperature range is generally regarded as indication of strange metals.
On the other hand, Fig.~\ref{fig:structure+rho}(d) demonstrates that the dc-resistivity of \CuBr\ and \CuNCS\ adopts the Fer\-mi-liquid behavior $\rho(T)\propto T^2$ before the onset of superconductivity ($T_c \approx$ 10 K).

\begin{figure*}
    \centering
        \includegraphics[width=0.8\linewidth]{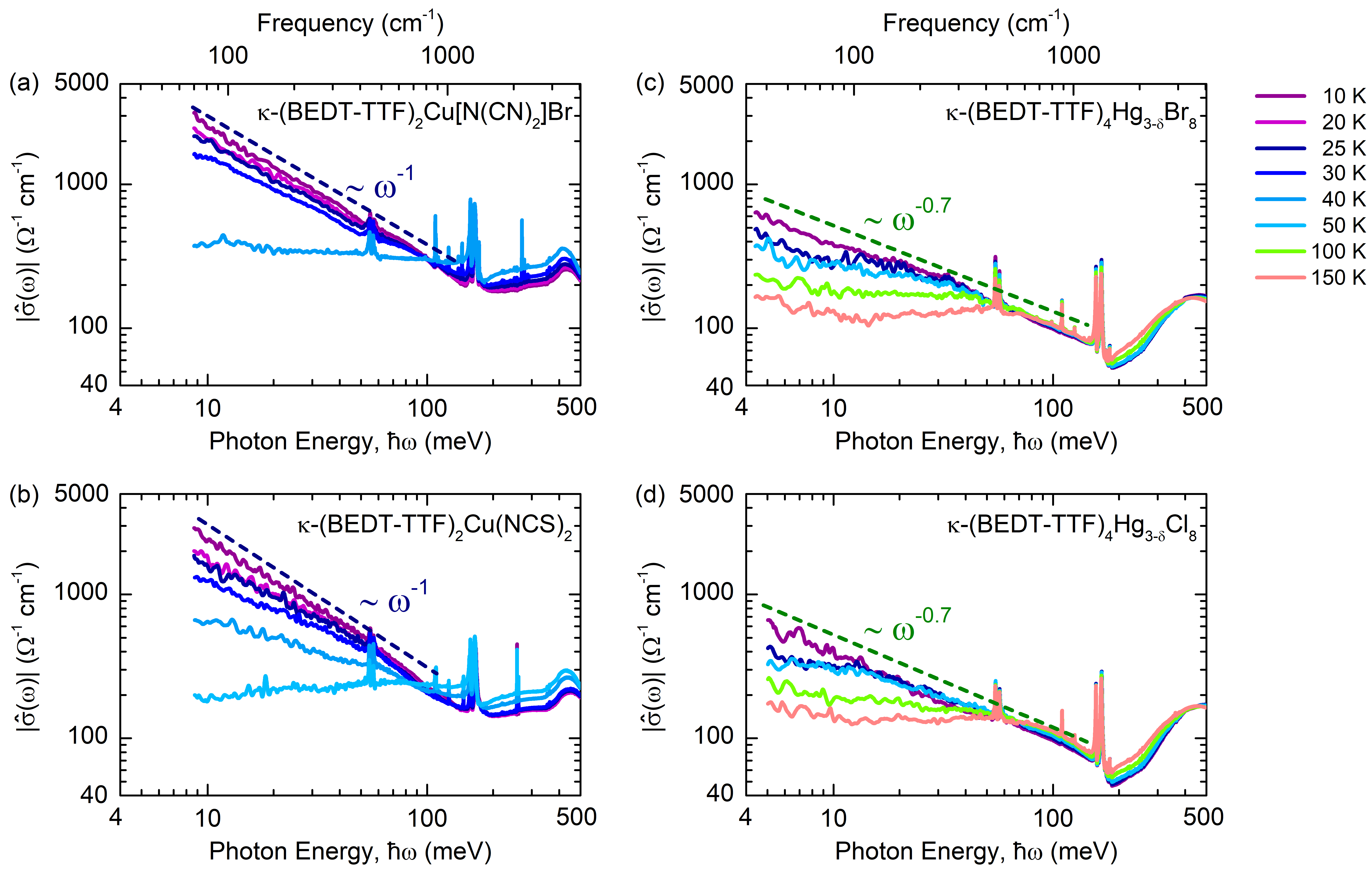}
        \caption{Absolute value of the complex conductivity as a function of frequency,
        $\vert \hat{\sigma}(\omega)\vert$, obtained at various temperatures as indicated. The free carrier response is observed predominantly up to 150~meV (1200~\cm), above which the electronic response is affected by Hubbard band and interband transitions, along with sharp molecular vibrations coupled to the electronic response.
        The  Drude-type $\omega^{-1}$ decay is observed for the Fer\-mi-liquid metals (a)~\CuBr\ and (b)~\CuNCS\ when  low temperatures are approached. For the strange metals (c)~\HgBr\ and (d)~\HgCl, the itinerant carriers follow a power-law $\vert \hat{\sigma}(\omega)\vert  \propto \omega^{-0.7}$, which is much flatter than that observed in the Fermi liquid compounds.
        All four compounds exhibit the in-plane structural and optical anisotropy typical for BEDT-TTF-based materials. The data shown here are for one crystal axis and the complete po\-lar\-iza\-tion-dependent results are presented in the Supplemental Material~\cite{SM}.}
    \label{fig:abs-sigma_w} 
\end{figure*}

Deeper insight into the microscopic mechanisms and the origin of the charge transport can be gained from the frequency-dependent complex  conductivity, $\hat{\sigma}(\omega)= \sigma_1(\omega) + {\rm i}\sigma_2(\omega)$~\cite{Dressel02Book}.
In Fig.~\ref{fig:abs-sigma_w}, a log-log presentation of the absolute value $\vert \hat{\sigma}(\omega) \vert$ is chosen; it offers a sensible comparison of optical properties as it takes both the absorptive ($\sigma_1$) and reactive ($\sigma_2$) components into account.
We focus on low temperatures to minimize the contributions from thermal excitations.
The comparison of $\vert \hat{\sigma}(\omega) \vert$ of the Fer\-mi-liquid compounds and strange metals evidences significant difference in charge-carrier relaxation.

Fig.~\ref{fig:abs-sigma_w}(a,b) illustrates that the correlated metals \CuBr\ and \CuNCS\ attain a linear decay $\vert \hat{\sigma}(\omega) \vert \sim \omega^{-1}$ in the range $T<40$~K,
where Fig.~\ref{fig:structure+rho}(d) proves Fer\-mi-liquid behavior by $\rho(T)\propto T^2$.
In other words, even for strong correlations the dynamics of the itinerant carriers is well captured by the Drude model at high frequency limit~\cite{Dressel02Book}:
\begin{equation}
\hat{\sigma} (\omega) =
\frac{\sigma_0}{1-{\rm i}\omega \tau} =
\frac{\omega_{\rm p}^2}{4\pi} \frac{1}{1/ \tau - {\rm i} \omega} \quad,
\label{eq:Drude}
\end{equation}
with the dc conductivity $\sigma_0 = N e^2 \tau / m$, the plasma frequency $\omega_p=\sqrt{4\pi N e^2/m}$ ($e$ is the elementary charge, $N$ the carrier density and $m$ their mass), and the relaxation time $\tau$.
However, for the \HgXd\ strange metals, the $\omega$-dependence strongly deviates from the conventional Drude roll-off. While the conductivity still follows a power law for almost two orders of magnitude in frequency, as seen in Fig.~\ref{fig:abs-sigma_w}(c,d),
it approaches a strongly reduced slope $\vert \hat{\sigma}(\omega) \vert \propto \omega^{-0.7}$ at low $T$. Note that this behavior resembles the power-law reported for high-$T_c$ cuprates \cite{Marel03, Michon23} and can be referred to as  ``stretched Drude behavior''~\cite{Fratini26}.

\begin{figure*}
    \centering
        \includegraphics[width=1\linewidth]{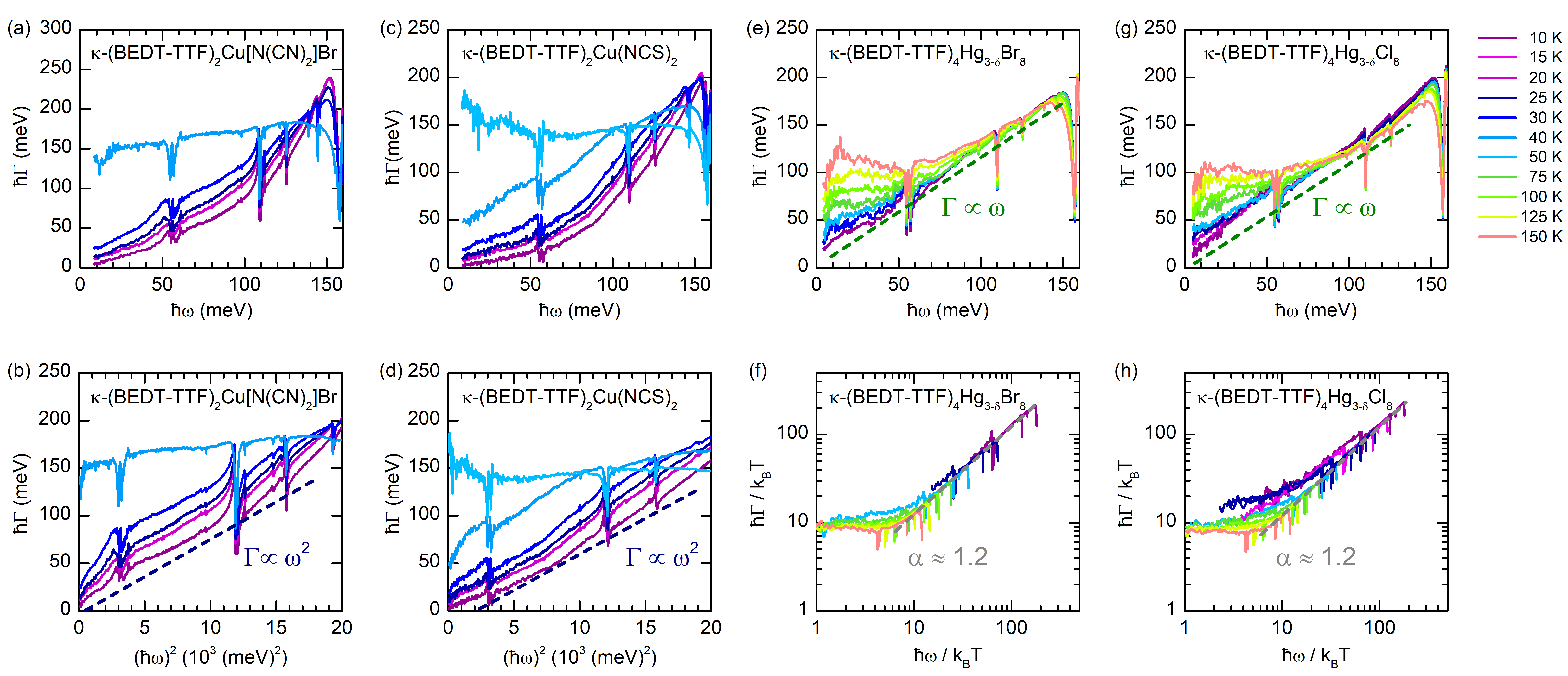}
        \caption{Frequency-dependent scattering rates extracted for different temperatures in the metallic states. The upper panels display $\Gamma(\omega)$ on a linear energy scale.
        The lower panels illustrate that the canonical metals (a,b) \CuBr\ and (c,d) \CuNCS\ approach a $\Gamma(\omega)\propto \omega^2$ behavior at 10~K expected for a Fermi-liquid response. In the case of the strange metals (e) \HgBr\ and (g) \HgCl, the scattering rate increases linearly with frequency: $\Gamma(\omega)\propto\omega$
        (the dashed lines are guides to the eye).
        Following the idea of Planckian scattering limit, in the two lower-row panels (f,h)
        the axes are scaled by the respective temperature. In a wide range, the data collapse to single curves. According to Eq.~(\ref{eq:Planckian}), we yield $\alpha \approx 1.2$ for both compounds \HgXd, as determined from a fit shown by the dashed lines.
         }
    \label{fig:gamma_w} 
\end{figure*}

To confirm the Fermi-liquid response of the itinerant carriers, the optical conductivity obtained from the Drude model of Eq.~(\ref{eq:Drude}) is customarily extended by introducing a frequency-dependent scattering rate~\cite{Dressel02Book}:
\begin{equation}
\Gamma(\omega) = \frac{\omega_{\rm p}^{2}}{4\pi} \frac{\sigma_{1}(\omega)}{\vert \hat{\sigma}(\omega)\vert^{2}}  \quad .
\label{eq:gamma-m}
\end{equation}
Fer\-mi-liquid theory predicts a quadratic frequency-de\-pen\-dent scattering rate, in addition to the $T^2$-dependence of the resistivity, summarized in the relation \cite{Berthod13,Maslov16,Pustogow21}:
\begin{equation}
\Gamma (T,\omega) \varpropto [ (2 \pi k_{B}T)^{2} + (\hbar\omega)^{2}]  \quad,
\label{eq:FermiLiq-gamma}
\end{equation}
where $k_{B}$ is the Boltzmann constant and $\hbar$ is the reduced Planck's constant. The temperature evolutions of this $\omega$-dependent scattering rate are compared in Fig.~\ref{fig:gamma_w}. The Fermi liquids \CuBr\ and  \CuNCS\ --~presented in panels (a) to (d)~-- approach $\Gamma(\omega) \propto \omega^2$  at low $T$, following Eq.~(\ref{eq:FermiLiq-gamma}).
Most importantly, the frequency-dependent scattering rate of the incommensurate compounds \HgXd\ significantly differs from that of the Fermi-liquid metals and acquires $\Gamma(\omega)\propto \omega$ increase over a large spectral range at low $T$, as illustrated in Fig.~\ref{fig:gamma_w}(e,g).
The clear deviations from the expected quadratic trend of $\Gamma(\omega)$
marks the breakdown of the conventional quasiparticle nature of carriers in these systems.

We remark that this linear-in-frequency response scales with universal Planckian dissipation, following
\begin{equation}
 \hbar\Gamma = \alpha k_{B}T  \quad ,
\label{eq:Planckian}
\end{equation}
where $\alpha$ is a sample-dependent prefactor on the order of unity~\cite{Patel19,Gleis25}. The ratio $\hbar \Gamma/k_BT$, which is plotted as a function of normalized frequency $\hbar\omega/k_BT$ in Fig.~\ref{fig:gamma_w}(f,h), approaches the scaling factor $\alpha \approx 1.2$ for both \HgBr\ and \HgCl. These numbers are consistent with slopes observed in other strange metals~\cite{Kimura25, Michon23}. The scaling applies up to about 150~meV (1200~\cm), i.e., the entire frequency range in which the optical response is dominated by itinerant carriers and relatively unaffected by interband transitions and molecular vibrations. The slight deviations observed for \HgCl\ are consistent with the recorded change in resistivity $\rho(T)$.

\textit{Discussion---}The family of $\kappa$-BEDT-TTF salts represents a well-established class of half-filled two-di\-men\-sion\-al electron systems where Coulomb repulsion determines the transport and electrodynamic properties~\cite{ToyotaBook,Powell11,Dressel20,Riedl22}.
While many compounds enter a Mott insulating state upon cooling,
\CuBr\ and \CuNCS\ are among the few examples which lie close to the Mott insulating state,
but exhibit metallic properties and then turn superconducting at ambient pressure.
Here, the resistivity follows a $T^2$-dependence and
the scattering rate increase as $\omega^2$;
according to Eq.~(\ref{eq:FermiLiq-gamma}), the materials are characterized as archetypal Fermi-liquid metals~\footnote{Our previous studies on the electrodynamics
of other $\kappa$-BEDT-TTF compounds came to similar conclusions~\cite{Dressel96,Dumm09,*Yasin11,Pustogow21}.}. The optical conductivity of the itinerant electrons yields a Drude-like decay $\vert \hat{\sigma}(\omega) \vert \propto \omega^{-1}$ up to 150~meV where other excitations occur.

Although \HgXd\ are highly-conducting molecular metals,
they never acquire conventional Fermi liquid signatures; instead they show all fingerprints of strange metals.
The distinct behavior observed in the transport and optical properties puts \HgXd\ outside a classification based solely on electronic correlations in half-filled Mott systems;
this discrepancy may be explained by structural considerations.
Guided by the discussion of the anomalous metallic state in high-$T_c$ cuprates,
it was proposed that the anomalous metallic response in \HgBr\
is a result of doping the BEDT-TTF layers beyond half filling due the non-stoichiometric Hg ions~\cite{Oike17,Oike24}, following early structural and stoichiometric arguments.
This fact, however, could not stand up to scrutiny in a recent spectroscopic study;
both \HgBr\ and \HgCl\ are similar to other $\kappa$-BEDT-TTF salts in terms of charge
per dimer estimated from molecular vibrations~\cite{Priya26}.

Since the conducting layers formed by dimerized BEDT-TTF molecules are very similar
in all $\kappa$-BEDT-TTF compounds, we need to take a close look at the incommensurate anion layers.
As illustrated in Fig.~\ref{fig:structure+rho}(a,b), the chain-like anions can be understood as a large-scale modulation of Hg atoms in midst of uniform columns formed by the Br/Cl atoms.
Our detailed single-crystal x-ray diffraction (XRD) study suggests similarities to incommensurate charge density waves: a large-scale modulation extending to few unit cells~\cite{Pouget24}.
This spacial modulation is limited only to the Hg atoms and it locally deforms the potential acting at the nearby  BEDT-TTF molecules. Although no structural defects are present, the effect mimics spatial disorder resulting in local variations in charge transfer.
This picture of an inhomogeneously varying chemical potential is supported by previous nuclear magnetic resonance (NMR) studies~\cite{Eto10}.

Suzuki {\it et al.} reported that with the application of hydrostatic pressure on \HgBr,  Fermi-liquid behavior is approached~\cite{Suzuki22}.
Since pressure and strain cannot change the stoichiometry determined by the synthesis,
it is more likely that compression suppresses the fluctuations induced by the spatial modulations:
the local potential becomes uniform and -- in addition -- the effective correlation strength decreases. This implies, in turn, that the presence of both strong correlation and spatial modulation are necessary to establish the strange metal behavior observed in \HgXd.

A closer look at Figs.~\ref{fig:abs-sigma_w}(a,b) reveals, that for increasing temperatures the Fermi-liquid behavior $\vert\hat{\sigma} (\omega)\vert \propto \omega^{-1}$ gradually shifts toward a strange metal with a reduced exponent of the power law. Similarly, we see from Fig.~\ref{fig:gamma_w}(a,c) that $\Gamma(\omega)$ approaches a linear trend for higher energies. These evolutions infer that, as the coherent motion of long-lived quasiparticle is reduced, the strange metal regime becomes apparent even at low energies. 

\textit{Conclusions---}Our comprehensive study establishes the molecular conductors \kHgXd\ as ``strange metals''
because (i) the dc resistivity rises as $\rho(T) \propto T$ to ambient temperature,
(ii) the complex conductivity reveals a stretched Drude decay $\vert\hat{\sigma} (\omega)\vert \propto \omega^{-0.7}$,
(iii) the scattering rate obtained from broad-range infrared measurements increases linearly with frequency, and (iv) follows the Planckian disspation.
These observations differ greatly from the Fermi-liquid state confirmed in organic superconductors \CuBr\ and \CuNCS\ by transport and optical properties: a quadratic increase of scattering rate with energy, and
a Drude-like drop of $\vert\hat{\sigma} (\omega)\vert\propto \omega^{-1}$.
The key difference lies in the crystal structure:
only regular anion layers allow for archetypal Fermi-liquid behavior at low temperatures.
The incommensurate compounds \HgXd\ possess
large-scale spatial modulations in the anion layer, which cause local chemical potential variations.
Our study implies that the structural modulation along with strong electronic correlations inherent to $\kappa$-BEDT-TTF salts leads to ``strange-metal'' features observed in the electrical and optical response.
Even though the Fermi-liquid state is generally one of the fundamental properties of strongly correlated metals, it is often suppressed by disorder, temperature etc. revealing the strange-metal behavior. Our results elucidate the link between the presence of randomness and the occurrence of the strange metal phase in strongly correlated electron systems.

\section*{Acknowledgments}
The authors acknowledge the technical support by Gabriele Untereiner. The authors acknowledge fruitful discussions with Matteo Mitrano, Natalia Drichko, and Jesse Liebman. The work at the University of Stuttgart was financially support by  Deutsche Forschungsgemeinschaft (DFG) Grants No. DR 228/68-1 and DR 228/74-1.

\nocite{Kini90, Urayama88,Hiramatsu15, Dolomanov09, Shibaeva88, Lyubovskaya85, Lyubovskaya87b, Li98, Sari23a, Sari24, Brown74, Pouget78, Volkova99, Chakraborty78, Gross86, Lyubovskaya87a, Takhirov90, Homes93, Tanner15, Eldridge91, Kornelsen91, Kanoda06, Faltermeier07}

\bibliography{ref}

\end{document}


\title{\text{Supplemental Material}\\
\textbf{Strange-metal behavior in correlated molecular conductors}}

\author{Savita Priya}
\email{savita.priya@pi1.uni-stuttgart.de}
\affiliation{1.~Physikalisches Institut, Universit\"{a}t Stuttgart, Pfaffenwaldring 57, 70569 Stuttgart, Germany}

\author{Maxim Wenzel}
\affiliation{1.~Physikalisches Institut, Universit\"{a}t Stuttgart, Pfaffenwaldring 57, 70569 Stuttgart, Germany}

\author{Christian Prange}
\affiliation{1.~Physikalisches Institut, Universit\"{a}t Stuttgart, Pfaffenwaldring 57, 70569 Stuttgart, Germany}

\author{Bertold Rasche}
\affiliation{Institut f{\"u}r Anorganische Chemie, Universit\"{a}t Stuttgart, Pfaffenwaldring 55, 70569 Stuttgart, Germany}

\author{Christian Sindlinger}
\affiliation{Institut f{\"u}r Anorganische Chemie, Universit\"{a}t Stuttgart, Pfaffenwaldring 55, 70569 Stuttgart, Germany}

\author{\\Takuya Kobayashi}
\affiliation{Graduate School of Science and Engineering, Saitama University, Saitama 338-8570, Japan}

\author{Hiromi Taniguchi}
\affiliation{Graduate School of Science and Engineering, Saitama University, Saitama 338-8570, Japan}

\author{Dita Puspita Sari}
\affiliation{Nuclear Structure Research Group, RIKEN Nishina Center, 
Saitama, 351-0198, Japan}

\affiliation{Shibaura Institute of Technology, 
Saitama 337-8570, Japan}

\author{Yasuyuki Ishii}
\affiliation{Shibaura Institute of Technology, 
Saitama 337-8570, Japan}

\author{\\Kazuya Miyagawa}
\affiliation{Department of Applied Physics, University of Tokyo, Tokyo 113-8656, Japan}

\author{Kazushi Kanoda}
\affiliation{1.~Physikalisches Institut, Universit\"{a}t Stuttgart, Pfaffenwaldring 57, 70569 Stuttgart, Germany}
\affiliation{Max Planck Institute for Solid State Research, Heisenbergstra\ss{}e 1, 70569 Stuttgart, Germany}
\affiliation{Department of Advanced Materials Science, University of Tokyo, Tokyo 113-8656, Japan}

\author{Simone Fratini}
\affiliation{Institut N\'{e}el - CNRS and Universit\'{e} Grenoble Alpes, 38042 Grenoble Cedex 9, France}

\author{Martin Dressel}
\email{dressel@pi1.physik.uni-stuttgart.de}
\affiliation{1.~Physikalisches Institut, Universit\"{a}t Stuttgart, Pfaffenwaldring 57, 70569 Stuttgart, Germany}
\date{\today}

\maketitle

\setcounter{secnumdepth}{3}
\setcounter{tocdepth}{3}
\tableofcontents
\clearpage

\section{Materials and Structure}

\subsection{Samples}
\begin{figure}[h!]
    \centering
        \includegraphics[width=0.35\linewidth]{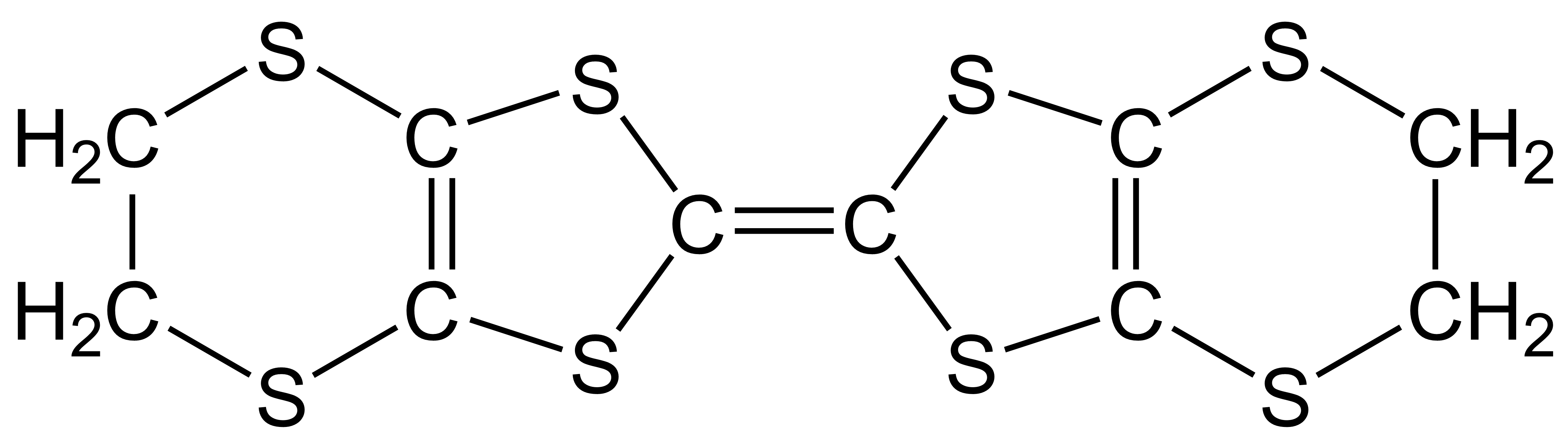}
        \caption{\textbf{BEDT-TTF molecule:} Chemical structure of the BEDT-TTF molecule, which forms the conducting layer of the materials under investigation.}
    \label{fig:ETmol}
\end{figure}

The electrochemical synthesis of \kCuBr{} and \kCuNCS{} single crystals (hereafter, abbreviated as \CuBr{} and \CuNCS, respectively, and together referred as Fermi-liquid compounds) was achieved as described in Refs.~\cite{Kini90, Urayama88}. Here BEDT-TTF stands for bis\-(ethylene\-di\-thio)-tetra\-thia\-fulvalene], C10~H8~S8; the chemical structure of the molecule is sketched in Fig.~\ref{fig:ETmol}. The crystalline samples are rhombus- or hexagonal-shaped with dimensions of about 1.5 $\times$ 2 mm$^{2}$ and thickness of 80-120 $\mu$m.

Crystals of the Hg-incommensurate charge transfer salts \kHgBrd{} (hereafter, abbreviated as \HgBr) and \kHgCld{} (hereafter, abbreviated as \HgCl) were synthesized by standard electrochemical methods described by
Lyubovskaya {\it et al.}~\cite{Lyubovskaya91}. The black single crystals are orthorhombic-shaped with flat surfaces, reaching a typical size of 1.5 $\times$ 1.7 mm$^{2}$ and thickness of 130-200 $\mu$m. The crystals are stable under ambient conditions but fragile. In general, they develop nice shiny as-grown surfaces, which are suited for optical reflection measurements without further treatment.
Our chemical analysis by XPS (X-ray photoelectron spectroscopy) confirmed a homogeneous composition of the atoms in anion layer throughout the samples. In particular, only one type of Hg chemical environment can be confirmed within the resolution limit.

\clearpage

\subsection{Crystal structure}
\subsubsection{Regular systems}
\label{sec:structureFermiliquidsystems}
\begin{figure}[h!]
    \centering
        \includegraphics[width=0.9\linewidth]{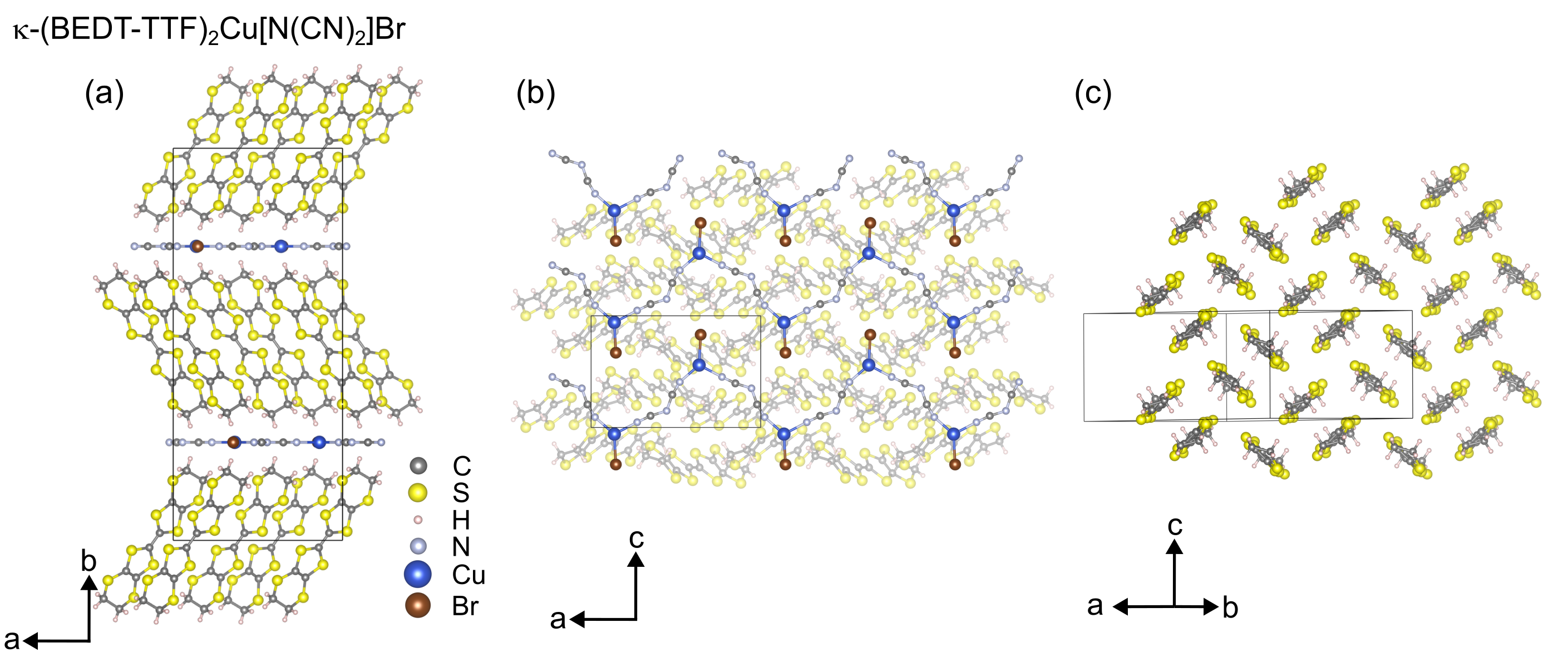}
        \caption{\textbf{Crystal structure of \kCuBr:} (a) Layered arrangement present in \CuBr, where dimerized organic BEDT-TTF molecules form the conducting layers; they alternate with the insulating sheets constituted by layers of polymeric Cu[N(CN)$_2$]Br units. (b) The relative arrangement of the conducting and insulating layers when viewed along the $b$-direction. (c) The orthogonal arrangement of BEDT-TTF dimers is the signature of the $\kappa$-phase~\cite{Hiramatsu15}.}
    \label{fig:CuBr-XRD}
\end{figure}

\begin{figure}[h!]
    \centering
        \includegraphics[width=0.9\linewidth]{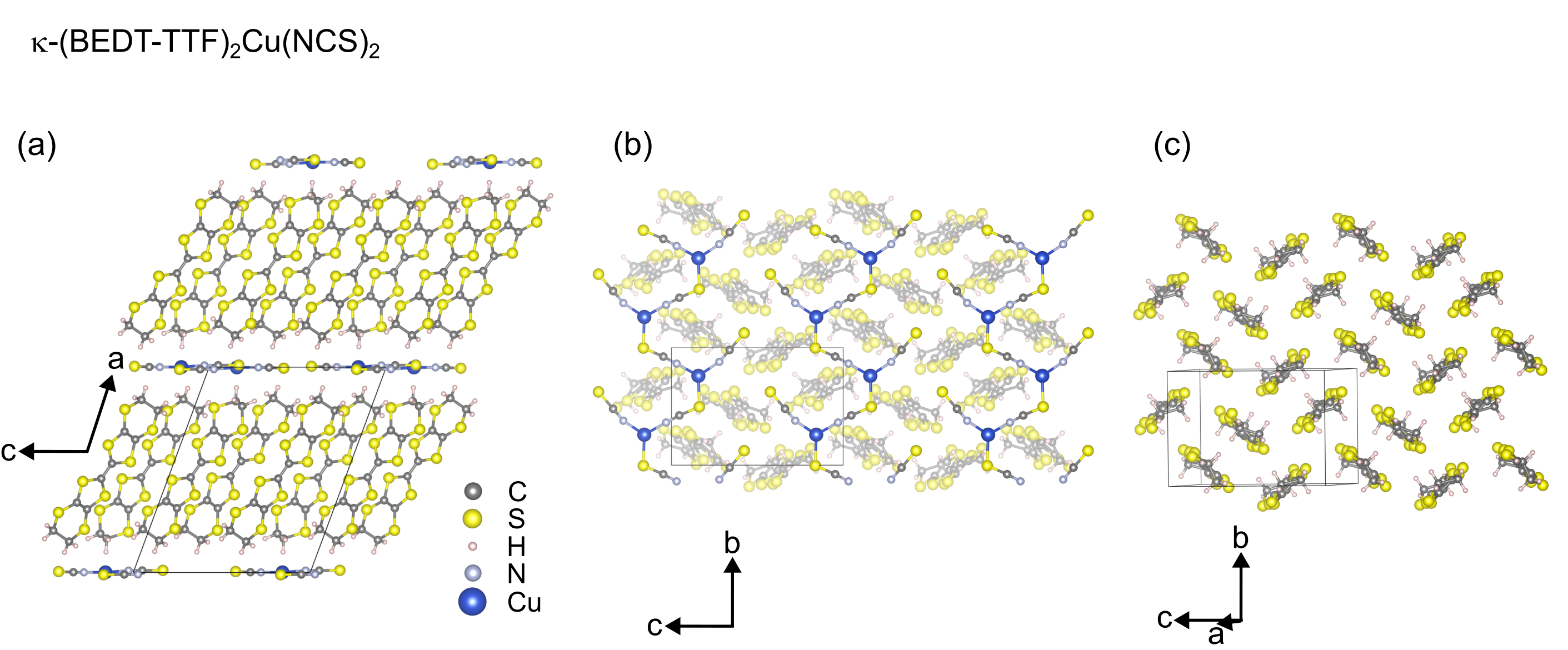}
        \caption{\textbf{Crystal structure of \kCuNCS:} (a) Layered arrangement present in \CuNCS, where the conducting layers are constituted by the organic BEDT-TTF molecules and the insulating layer is formed by Cu(NCS)$_2$ layers. (b) The relative arrangement of the conducting and insulating layer when looking at the $bc$-plane. (c) The signature $\kappa$-arrangement is observed in the conducting layer \cite{Hiramatsu15}.}
    \label{fig:CuNCS-XRD}
\end{figure}

The crystal structures of the renowned organic superconductors \CuBr\ (CCDC No.: 1016184) \cite{Hiramatsu15} and \CuNCS\ (CCDC No.: 1016179) \cite{Hiramatsu15} are obtained from the crystallographic database and presented in Figs.~\ref{fig:CuBr-XRD} and \ref{fig:CuNCS-XRD}.
The unit cell parameters are listed in Tab.~\ref{tab:unitcell}.

\label{sec:structureFermiliquids}

\clearpage

\subsubsection{Hg-incommensurate systems}
\label{sec:structureHgsystems}

Single crystal x-ray diffraction (XRD) measurements were performed for \kHgBrd\ and \kHgCld,  utilizing a Rigaku XtaLAB Synergy diffractomer (Source: Cu K$_{\alpha}$) at University of Stuttgart. The data were collected and processed by the CrysAlis software. Well shaped single crystals with approximately $50 \times 50~\mu{\rm m}^2$ were selected for x-ray measurements.
The samples can be twinned, which was taken into account during the structural analysis. The electron density was modeled, processed and visualised by the OLEX2 software \cite{Dolomanov09}. The electron density for Hg-atoms was analysed as fractional occupancy at sites with high-occupancy probability, estimated from electron density.
The procedure is similar to the first report by Shibaeva and Rozenberg~\cite{Shibaeva88} and Lyuboskaya {\it et al.} \cite{Lyubovskaya85,Lyubovskaya87b}. Alternatively, the structure could be resolved by treating the Hg atoms as a different sublattice, as performed by Naughton and collaborators \cite{Li98}. However, the latter method can limit the analysed structure of the Hg atoms to a small number of unit cells, which leads to inaccurate positions of Hg atoms when the structure is expanded.
To reduce the crystal structure to a unit cell comparable to other BEDT-TTF salts, we treat the Hg atomic positions as fractional occupancy sites.

\begin{figure}[h!]
    \centering
        \includegraphics[width=0.88\linewidth]{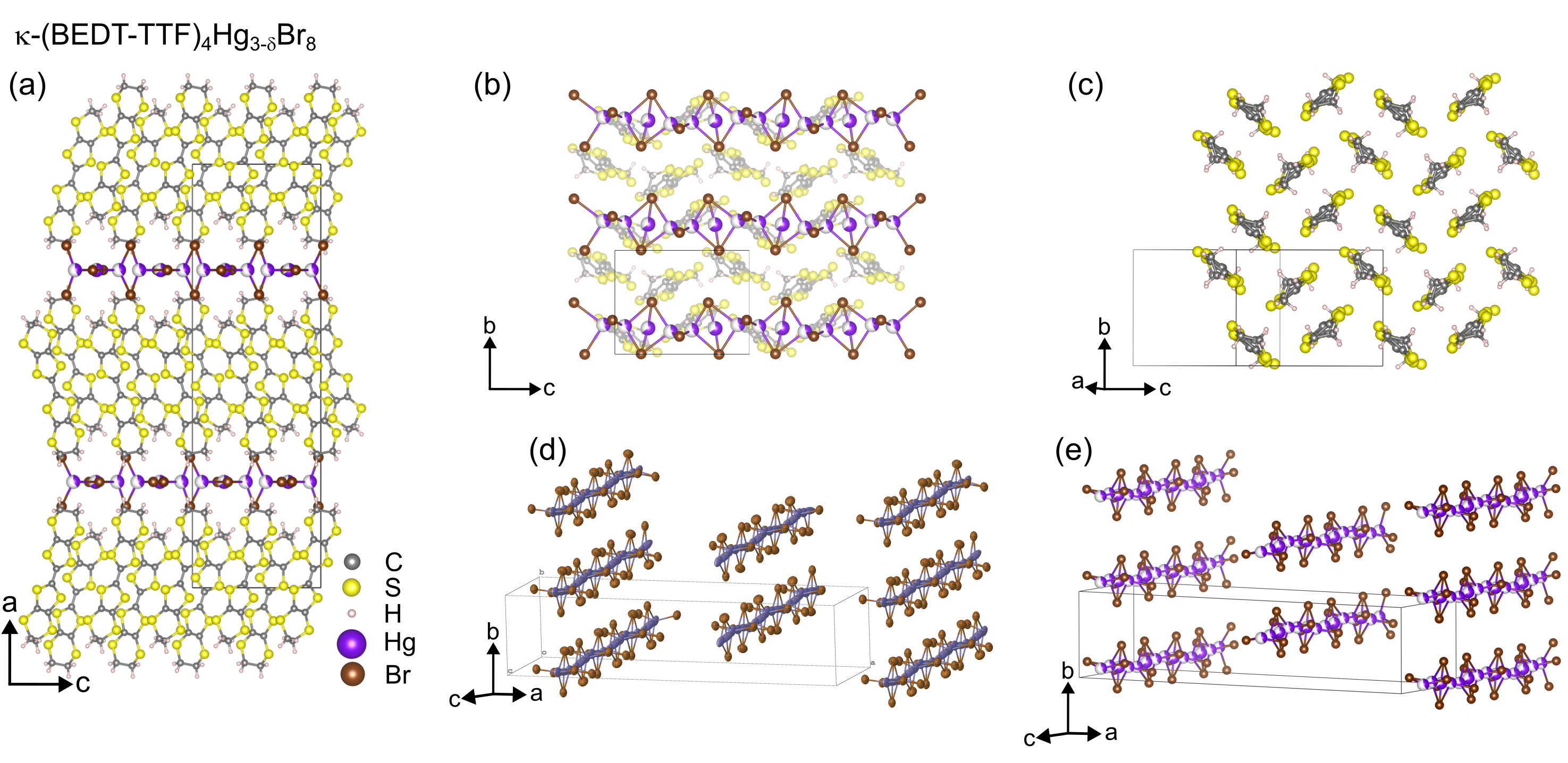}
        \caption{\textbf{Crystal structure of \kHgBrd:} (a) Layered arrangement present in \HgBr, where the conducting layer is formed by organic BEDT-TTF molecules and the insulating layer is formed by Hg-Br layer. (b) The relative arrangement of the conducting and insulating layer along $bc$-plane. (c) The signature $\kappa$-arrangement observed in the conducting layer. (d) The electron density plot obtained for the anion Hg-Br layer and (e) the modeled Hg atomic positions, with exact value of Hg$_{2.90}$Br$_8$, where the incomplete spheres denoted the averaged fractional occupancies.}
    \label{fig:HgBr-XRD}
\end{figure}

\begin{figure}[h!]
    \centering
        \includegraphics[width=0.88\linewidth]{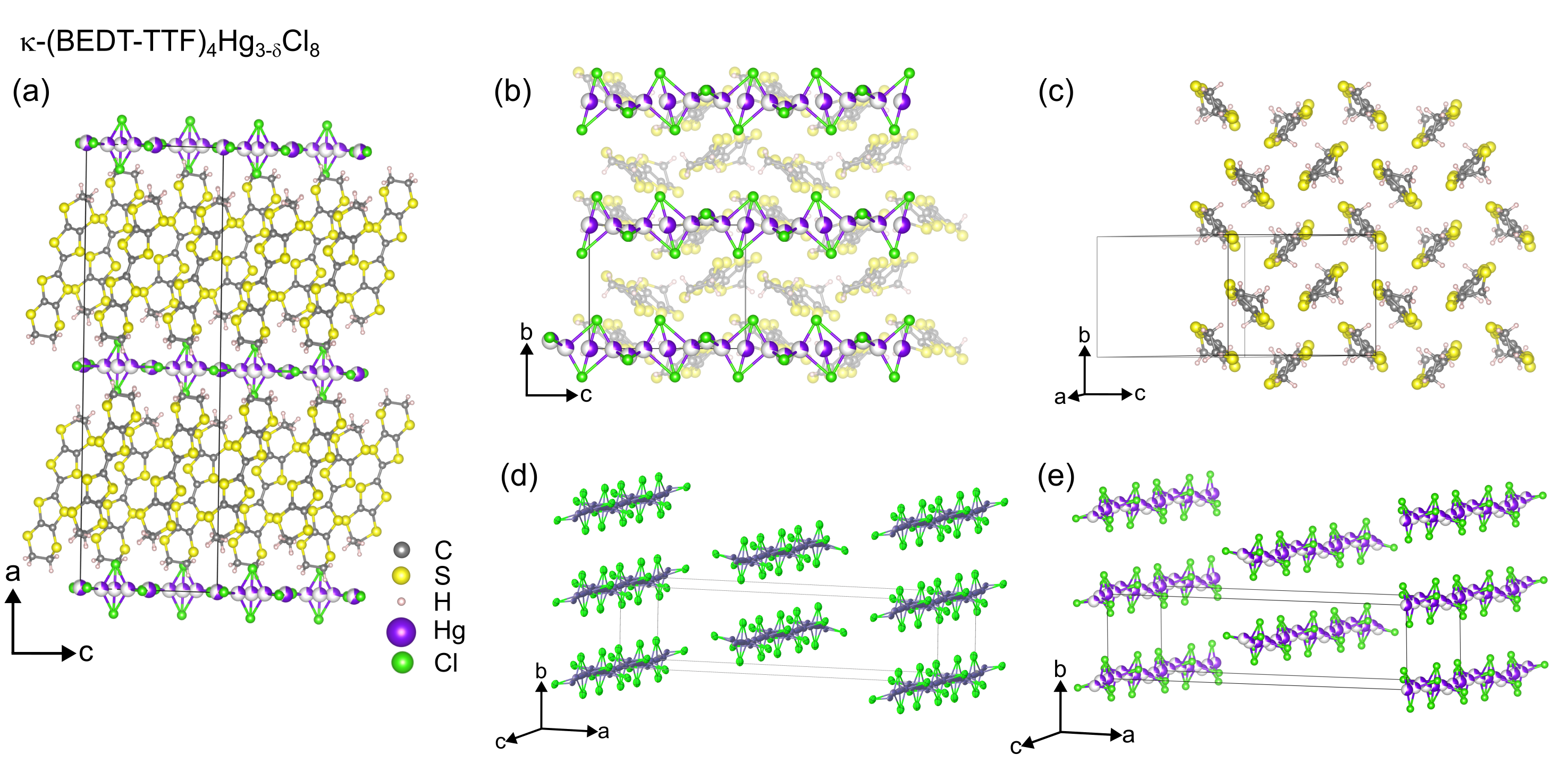}
        \caption{\textbf{Crystal structure of \kHgCld:} (a) Layered arrangement present in \HgCl, where the conducting layer is formed by organic BEDT-TTF molecules and the insulating layer is formed by Hg-Cl layer. (b) The relative arrangement of the conducting and insulating layer along $bc$-plane. (c) The signature $\kappa$-arrangement observed in the conducting layer. (d) The electron density plot obtained for the anion Hg-Br layer and (e) the modeled Hg atomic positions, with exact value of Hg$_{2.82}$Cl$_8$, where the incomplete spheres denoted the averaged fractional occupancies.}
    \label{fig:HgCl-XRD}
\end{figure}

From our room-temperature XRD investigations we obtain the space group \textit{Pmcn} for \HgBr\ and \textit{I}2\textit{/c} for the \HgCl\ crystals.
The resulting structures of the incommensurate compounds \HgBr\ and \HgCl\ are presented in Figs. \ref{fig:HgBr-XRD} and \ref{fig:HgCl-XRD}.
The respective unit cell parameters are listed in Tab.~\ref{tab:unitcell}.

\begin{table}[h]
\centering
\caption{Crystallographic parameters of the unit cell for the investigated compounds
\HgBr, \HgCl, \CuBr\ and \CuNCS, recorded at ambient conditions. }
\label{tab:unitcell}
\begin{tabular}{l|c|c|c|c|c|c|c|c|c}
\toprule
Compound & Space  & \multicolumn{6}{c|}{Lattice parameters} & Unit cell & Ref.\\
&group & \multicolumn{6}{c|}{} & volume & \\
&			& 	$a$ (\AA) & $b$ (\AA) & $c$ (\AA) & $\alpha$ (\textdegree) & $\beta$ (\textdegree) & $\gamma$ (\textdegree) & (\AA$^3$)	\\
\midrule
$\kappa$-(BEDT-TTF)$_4$Hg$_{3-\delta}$Br$_8$  	& \textit{Pmcn} (62) &	37.1522(7) & ~8.7052(3)	& 11.2571(3)	&	90.000 &	~90.000 & 90.000 & 3640.74(15) & This work	\\
$\kappa$-(BEDT-TTF)$_4$Hg$_{3-\delta}$Cl$_8$ 		& \textit{I}2\textit{/c} (15) & 35.9490(6)	&	~8.7502(1) &	11.0654(2)  &	90.000 &	~91.040(3) &	90.000 & 3480.17(9) & This work	\\
$\kappa$-(BEDT-TTF)$_2$Cu[N(CN)$_2$]Br 		& \textit{Pnma} (62) & 12.9653	&	30.0280 &	~8.5444  &	90.000 & ~90.000 &	90.000 & 3326.523109 & \cite{Hiramatsu15} 	\\

$\kappa$-(BEDT-TTF)$_2$Cu(NCS)$_2$ 		& \textit{P}2\textit{$_1$} (4) & 16.2397	&	~8.4425 &	13.1293  &	90.000 &	110.276 &	90.000 & 1688.532098 &	\cite{Hiramatsu15} \\
\bottomrule
\end{tabular}
\end{table}

The crystallographic structure of \kHgXd\ contains layers of BEDT-TTF dimers in the typical
$\kappa$-phase arrangement displayed in Figs.~\ref{fig:HgBr-XRD}(c) and \ref{fig:HgCl-XRD}(c).
The dimers are rotated approximately $90^{\circ}$ with respect to each other and are slightly tilted in $c$-direction.
While \HgCl\ crystallizes in a monoclinic structure (\textit{I}2\textit{/c}) with all BEDT-TTF dimers are tilted in the same direction (parallel), for our \HgBr\ crystals we identify an orthorhombic unit cell
(space group \textit{Pmcn}) where the BEDT-TTF dimers are tilted in alternating directions (zig-zag sequence of adjacent BEDT-TTF layers along  $a$-direction) as shown in Fig.~\ref{fig:HgBr-XRD}.
The observed x-ray pattern is well distinct from the one obtained for
twinned monoclinic crystals.

A monoclinic arrangement was previously suggested for both compounds  \cite{Shibaeva88,Lyubovskaya87b,Sari23a,Sari24}.
Li \textit{et al.} reported monoclinic and orthorhombic modifications of \HgBr, with superconductivity linked to the monoclinic phase \cite{Li98}.
However, we also observed superconductivity for crystals of the orthorhombic phase.
Although a systematic study on this topic has yet to be conducted, we believe that these variations do not affect the physics observed in these materials.
The electronic properties are confined to the BEDT-TTF layers which are essentially the same in both configurations.
This conclusion is confirmed by the comparison of the Fermi-liquid metals and superconductors
\CuBr\ and \CuNCS\ presented in Sec.~\ref{sec:structureFermiliquidsystems}.
\CuBr\ crystallizes in the orthorhombic phase with zig-zag arrangement of BEDT-TTF dimers in adjacent layers, as shown in Fig.~\ref{fig:CuBr-XRD}(a), while the unit cell of
\CuNCS\ [Fig.~\ref{fig:CuNCS-XRD}(a)] is monoclinic with parallel BEDT-TTF orientation of the dimers.
Nevertheless, the electronic and optical properties observed in these compounds are very similar.

The BEDT-TTF layers in \HgXd\ are separated by anion layers containing halides, Br or Cl, and Hg atoms.
When analyzing our XRD results assuming fractional occupancies of the Hg atoms, we obtain
the unit cell formula:
\begin{itemize}
\item For \HgBr: C40 H32 Br8 Hg2.90 S32, with 2.90 Hg atoms per unit cell; corresponding to $\delta = 0.10\pm 0.02$.
\item
For \HgCl: C40 H32 Cl8 Hg2.82 S32, with 2.82 Hg atoms per unit cell; corresponding to $\delta = 0.18 \pm 0.02$.
\end{itemize}
The amount of Hg atoms is obtained by modeling the electron density in the columns of halide chains, extending along the $c$-direction.
In order to account for the uncertainty and model dependence in the structural analysis of the incommensurate compounds, we have used the notation of Hg$_{3-\delta}$ throughout the paper.

As illustrated in Figs.~\ref{fig:HgBr-XRD} and \ref{fig:HgCl-XRD} the crystal structure of \HgXd\ can be pictured as follows:
the halide ions are linked to the BEDT-TTF molecules, observing a fixed stoichiometry of 2:1.
The halides form slightly distorted tetrahedra, which share common apices, edges, and faces forming an infinite zig-zag chain stretching along the $c$-axis.
This scaffold is inhomogeneously decorated by Hg ions, resulting in an incommensurate structure.
As a matter of fact,
the incommensurate arrangement of Hg ions along chains in an ionic scaffold is well known with particular emphasis on the interatomic distance within the polyatomic chains as well as the bond length to neighboring ions \cite{Brown74,Pouget78,Volkova99}. The characterization of the superconducting state in the incommensurate mercury chains of Hg$_{2.86}$AsF$_6$ have been controversially discussed for many years \cite{Chakraborty78,Gross86}.
In the series of \HgXd, being examined here, we find that the modulation due to incommensuration diminishes with increasing size of the halide ions.

Our results are consistent with values reported by the Chernogolovka group: $\delta = 0.22$ in the case of \HgCl\ and $\delta = 0.11$ in the case of \HgBr; for the iodine compound $\delta = 0$  \cite{Lyubovskaya87a,Shibaeva88,Takhirov90,Lyubovskaya91}.
The detailed crystallographic investigations on \HgBr\ by the SUNY group at Buffalo also aligns with these conclusions \cite{Li98}. They consider the ratio of the unit cell volume
of the BEDT-TTF and Br sublattice (I) with respect to the unit cell volume of the Hg sublattice (II): $V^{I}/V^{II} =2.883$. At a reduced temperature of $T=40$~K, Li {\it et al.} found a triclinic phase with $V^{I}/V^{II} =2.868$ \cite{Li98}.


\section{Transport measurements}

Temperature-dependent transport measurement were performed by the standard four-contact method, utilizing a Physical Property Measurement System (PPMS) of Quantum Design. The four electrical contacts were placed on a straight line, along the two in-plane crystallographic axes (identified by infrared reflection measurements) on the flat surface of the single crystal; thin gold wires were attached to the sample with the help of carbon paste.

\begin{figure}[h!]
    \centering
        \includegraphics[width=0.75\linewidth]{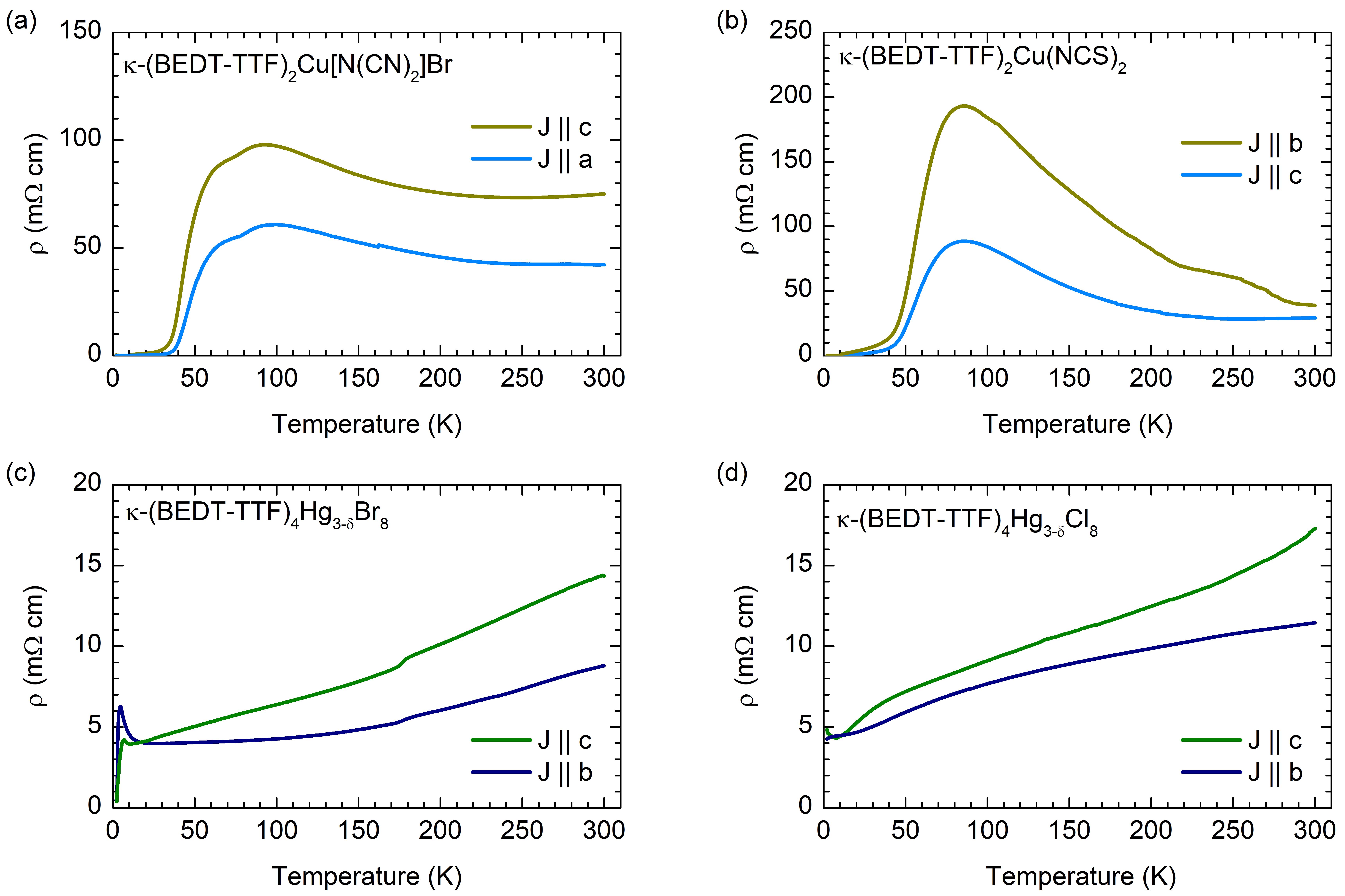}
        \caption{\textbf{Temperature dependent dc resistivity:} The in-plane electrical resistivity was measured along two crystallographic directions. The two Fermi-liquid compounds (a) \CuBr\ and (b) \CuNCS\ start with room temperature values of approximately 50~m$\Omega$cm. Metallic resistivity is observed only below approximately 80~K.
        The strange metal compounds (c) \HgBr\ and (d) \HgCl\ show a much lower room-temperature resistivity of $\rho \approx 10-15~{\rm m}\Omega$cm, with a similar anisotropy. Most important is the continuous decrease $\rho(T) \propto T$ in an extended temperature range.}
    \label{fig:resistivity}
\end{figure}

Fig.~\ref{fig:resistivity}(a,b) display the dc resistivity curves of the Fermi-liquid metals
\CuBr\ and \CuNCS, which become superconducting at about 10~K.
Upon cooling, they exhibit a maximum in resistance around $T=80-100$~K. Only below approximately 40~K, $\rho(T)\propto T^2$ is observed. The BEDT-TTF arrangement in all these $\kappa$-phase salts induces crystallographic anisotropy, which is also evident from the slightly lower resistance along the direction the BEDT-TTF dimers are tilted; which is $J \parallel a$ for \CuBr\ and $J \parallel c$ for \CuNCS. However, the anisotropy is limited to quantitative behavior and small details/ features, while the qualitative response remains analogous within the conducting plane.

In the lower panels of Fig.~\ref{fig:resistivity}, the temperature dependence of the resistivity is shown for the strange metals (c) \HgBr\ and (d) \HgCl. It continuously drops with reducing the temperature. Shortly before superconductivity sets in, a minor maximum is observed below 10~K.
Over an extended temperature range the data can be well described by $\rho(T)\propto T$; no indications of a Fermi-liquid behavior.

\clearpage

\section{Optical properties}
\subsection{Infrared spectroscopy}

In-plane temperature-dependent infrared measurements were performed on the large flat surface of the samples; suitable polarizers allow for directional dependent results. The reflectivity in the spectral range of 600-18\,000~\cm\ was obtained via a Bruker Vertex 80v spectrometer coupled with a Hyperion microscope and a CryoVac Konti micro helium-flow cryostat, utilizing a freshly evaporated gold mirror as reference. The reflectivity measurements in the frequency region of 35-600 \cm{} were carried out by a Bruker IFS-113v spectrometer with a custom-built helium cryostat. The absolute reflectivity value was obtained by \textit{in-situ} gold-overcoating in the far-infrared region \cite{Homes93}; the reflectivity spectra of other regions were normalized with respect to the absolute reflectivity value.

Towards low frequencies, the data were extrapolated using the Hagen-Rubens relation $R(\omega) = 1 - (2\omega/\pi\sigma_{dc})^{1/2}$, when the reflectivity approaches the absolute values of close to unity. This is the case in the complete temperature range for \HgBr, \HgCl. For \CuBr{} and \CuNCS{}
the Hagen-Rubens extrapolation applies at temperatures below 50~K, where metallic transport is observed in dc resistivity. For the high temperature spectra of \CuBr{} and \CuNCS, the absolute values of reflectivity are around 0.6, and an exponential relation is applied for extrapolation similar to the Hagen-Ruben relation: $R(\omega) = 1 - (2\omega/\pi\sigma_{dc})^{1/x} (x=4-6)$; this allows us to match the experimental reflectivity values rather well. The data above 18\,000~\cm\ were extrapolated using the x-ray scattering functions~\cite{Tanner15}. Note, the extrapolation is necessary for the mathematical Kramers-Kronig transformation but does not affect the optical conductivity in the experimentally measured energy range~\cite{Dressel02Book}. The experiments were performed at fixed temperatures from $T=300$~K down to 10~K.

The experimentally determined reflectivity $R(\omega)$ was utilised to calculate the phase $\theta$($\omega$) by Kramers-Kronig transformation and subsequently the complex dielectric response $\hat{\varepsilon}(\omega) = \varepsilon_1(\omega) + {\rm i}\varepsilon_2(\omega)$.
The complex optical conductivity $\hat{\sigma}(\omega)  = \sigma_{1}(\omega) + {\mathrm i}\sigma_{2}(\omega)$ is obtained from the complex dielectric response utilizing:
\begin{equation}
\hat{\sigma}(\omega)=  - \frac{i \omega}{4 \pi} [\hat{\varepsilon} (\omega) - \varepsilon_{\infty}] \quad ,
\label{eq:sigmacomplex}
\end{equation}
where $\sigma_1(\omega)$ and $\sigma_2(\omega)$ are real and imaginary parts of the complex conductivity, $\varepsilon_0$ is the vacuum permittivity, and $\varepsilon_{\infty}$ is the high-frequency dielectric constant.
The absolute value of the optical conductivity $\vert\hat{\sigma}(\omega)\vert$ and the phase angle of the conductivity $\phi$ are given by:
\begin{equation}
\vert\hat{\sigma}(\omega)\vert = \sqrt{\sigma_{1}(\omega)^{2} + \sigma_{2}(\omega)^{2}}
\quad {\rm and} \quad
\phi(\omega) = \mathrm{arctan} \left\lbrace \frac{\sigma_2 (\omega)}{\sigma_1 (\omega)} \right\rbrace  \quad.
\end{equation}
We refer to Ref.~\cite{Dressel02Book} for more details on the optical conductivity relations.

The frequency-dependent optical response is commonly expressed by the Drude-Sommerfeld model with a fixed but temperature dependent scattering rate  $1/\tau$:
\begin{equation}
\hat{\sigma} (\omega) = \frac{\omega_{\mathrm{p}}^2}{4\pi} \frac{1}{1/\tau - i \omega}  \quad .
\label{eq:Drude-Sommerfeld}
\end{equation}
The Eq.~(\ref{eq:Drude-Sommerfeld}) gives the $\vert\hat{\sigma}(\omega)\vert \propto \omega$ dependence at high frequencies, often described as Drude roll-off.

\clearpage
\subsection{Temperature-dependent optical response}

\begin{figure}[h!]
    \centering
        \includegraphics[width=0.61\linewidth]{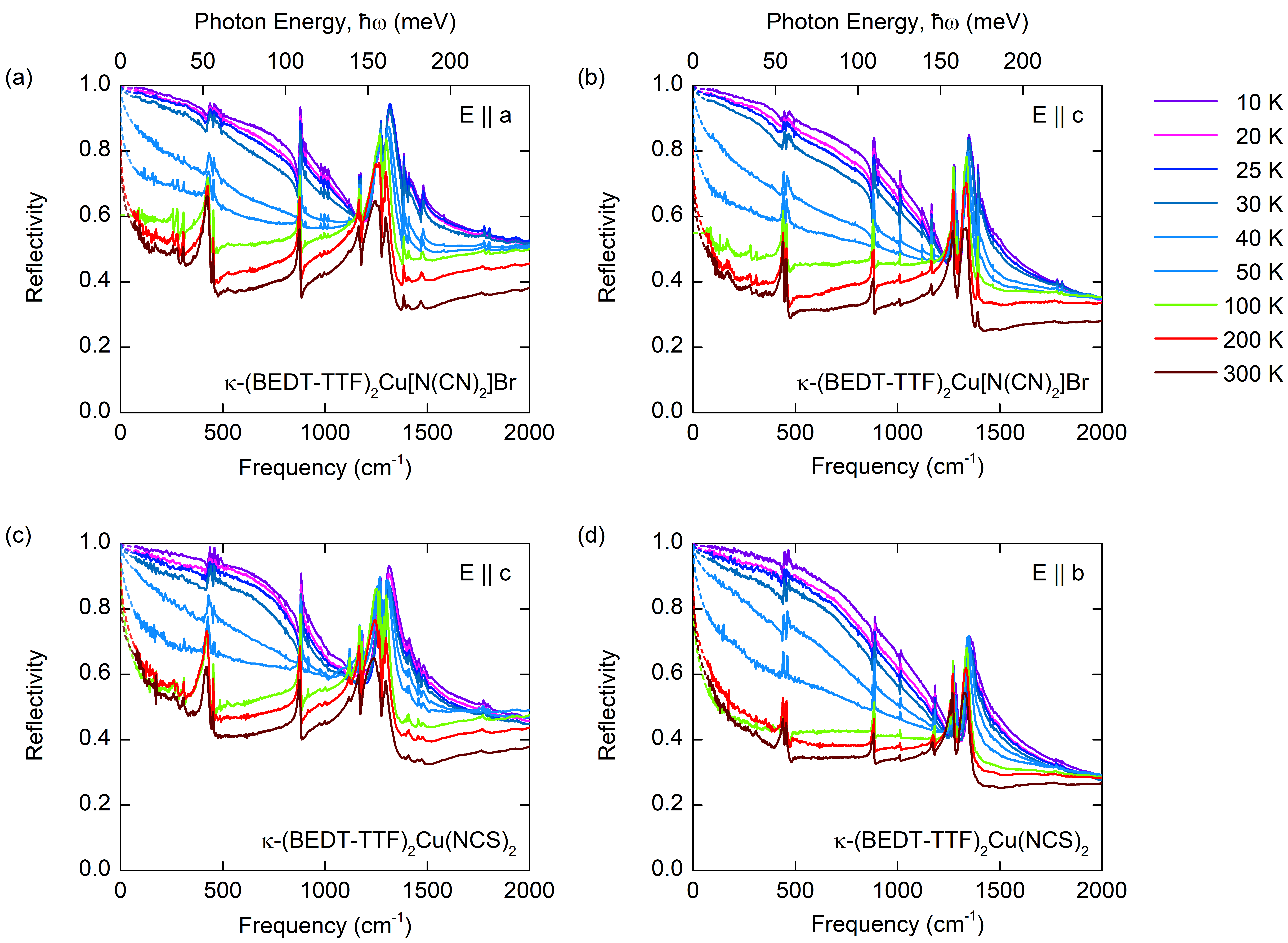}
        \caption{\textbf{Reflectivity of Fermi-liquid compounds at different polarization directions:} The experimentally obtained in-plane optical reflectivity, $R(\omega)$, for the Fermi-liquid compounds: (a,b) \CuBr, and (c,d) \CuNCS. The data are presented in linear scale for low energies, where signatures of the plasma edge can be identified around 700~\cm.}
    \label{fig:Ref-lin-FL}
\end{figure}

\begin{figure}[h!]
    \centering
        \includegraphics[width=0.61\linewidth]{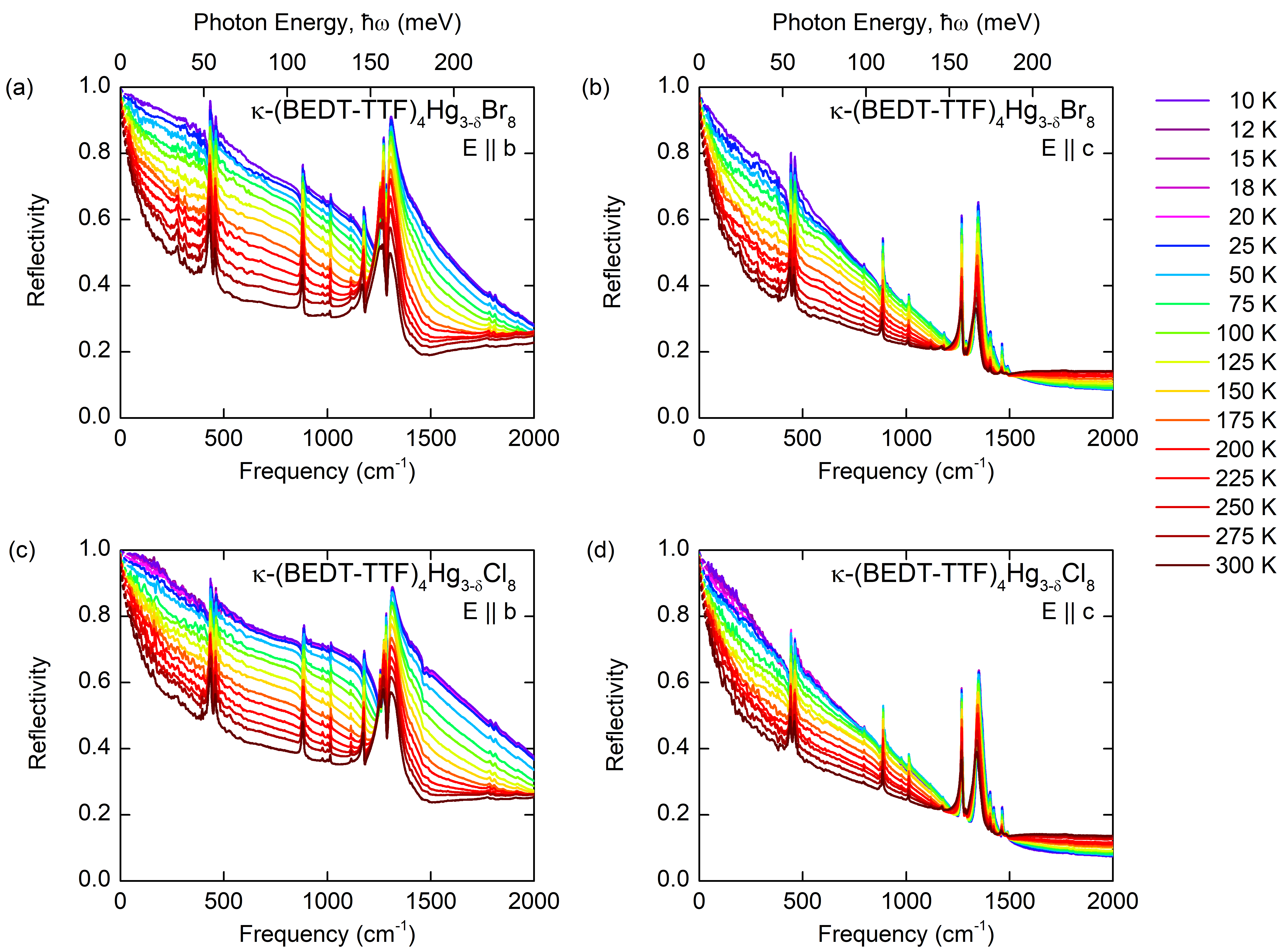}
        \caption{\textbf{Temperature evolution of the polarization-dependent reflectivity of strange-metal compounds:} The experimentally obtained in-plane optical reflectivity, $R(\omega)$, for the strange-metal compounds: (a,b) \HgBr, and (c,d) \HgCl. The data are presented in linear scale for low energies, where clear signatures of the plasma edge are absent and at low temperatures, almost linear dependence with energy can be realised.}
    \label{fig:Ref-lin-SM}
\end{figure}

Figs. \ref{fig:Ref-lin-FL} and \ref{fig:Ref-lin-SM} presents the experimentally obtained reflectivity, $R(\omega)$, for the Fermi-liquid and strange-metal compounds, magnified at low energies.
The differences between the Fermi-liquid and strange-metal compounds are evident from their raw $R(\omega)$ spectra, where the former shows signatures of a plasma edge, centered around 700~\cm, while the latter illustrates almost linear dependence with energy, in the low-temperature limit.

\clearpage

\subsubsection{Fermi liquids}
In Fig.~\ref{fig:CuBr-Ref} we plot the temperature evolution of the optical properties of \CuBr\ measured in a wide spectral range parallel the $a$ and $c$-axes. Note the logarithmic frequency scale.
Corresponding to the two-dimensional electronic properties, the overall behavior is rather similar for both polarizations. Interband excitations are the main cause of anisotropy in the mid-infrared range.
As the temperature is lowered, the infrared reflectivity $R(\omega)$ below 1000~\cm\ develops a pronounced plasma edge for both polarizations. This corresponds to the Drude-like response of the itinerant carriers that gains spectral weight below 50~K (panels c,d). This is about the temperature, when the dielectric constant crosses to negative values, as shown in the lower panels (g,h).
While the properties in the far-infarred range are dominated by the itinerant carriers,
the optical response above 1250~\cm, {\it i.e.} around 150~meV, is dominated by molecular vibrations and interband tranistions.
The mid-infrared maximum is due to interband transitions and excitations between the Hubbard bands.
It causes the rapid drop of $R(\omega)$ at 5000~\cm.

The results for \CuBr{} are consistent with the previously reported temperature-dependent reflectivity, and conductivity data \cite{Eldridge91,Dumm09}, revealing in-plane anisotropic spectra and anisotropic temperature-dependent evolution of interband transitions.

\begin{figure}[h!]
    \centering
        \includegraphics[width=0.73\linewidth]{SI-CuBr-Ref-S1-S2.png}
        \caption{\textbf{Optical response of \kCuBr:} (a,b) Optical reflectivity, $R(\omega)$, of \CuBr\ measured in the temperatures range between 10 and 300 K for two in-plane polarization axes, $E \parallel a$ and $E \parallel c$. (c,d) The real part of optical conductivity $\sigma_1(\omega)$ and (e,f) the  imaginary part $\sigma_2(\omega)$. (g,h) Temperature evolution of the respective dielectric constant $\varepsilon_1(\omega)$. The low-energy extrapolation in the reflectivity (below 70 \cm) in panels (a,b) is indicated by dashed lines. Below $T=50$~K, \CuBr{} exhibits drastic changes in the optical response, below 3000~\cm, as is evident from sharp increase in the values of $\sigma_1$ and $\sigma_2$, and observation of zero-crossing around 2000-3000 \cm{} in $\varepsilon_1$, marking the metallic response below 50 K for \CuBr.}
    \label{fig:CuBr-Ref}
\end{figure}

\clearpage

\begin{figure}[h!]
    \centering
        \includegraphics[width=0.8\linewidth]{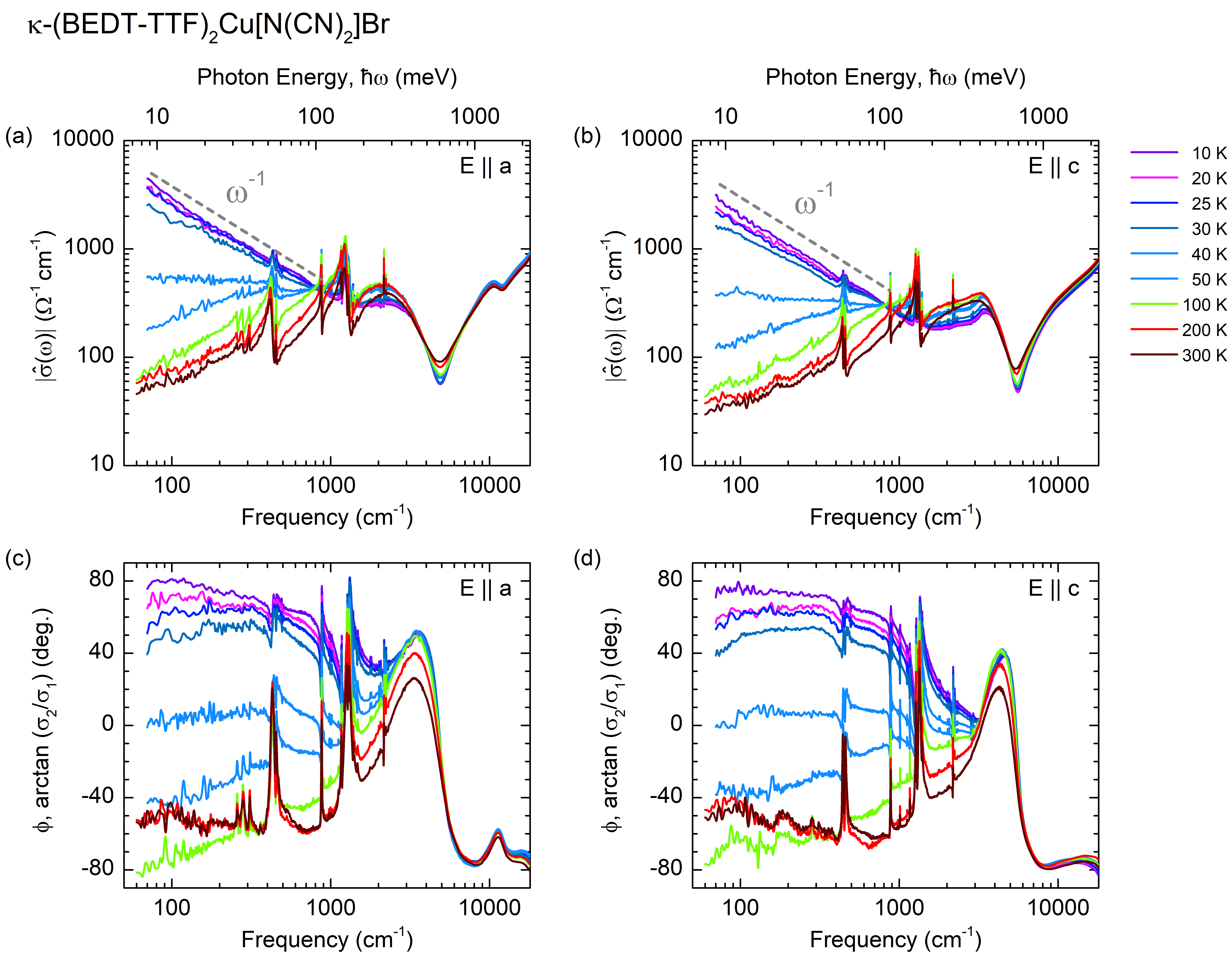}
        \caption{\textbf{Absolute value of optical conductivity, $\vert\hat{\sigma}(\omega)\vert$, and phase angle, $\phi(\omega)$, of \kCuBr:} (a,b) Temperature-dependent absolute value of optical conductivity, for the two in-plane polarization axes ($E \parallel a$ and $E \parallel c$) for \CuBr, illustrates that at low temperatures and low frequencies, $\vert\hat{\sigma}(\omega)\vert \propto \omega^{-1}$; and (c,d) the phase angle, $\phi(\omega)$, values get constant and obtain values to 80$^\circ$ at low temperatures.}
    \label{fig:CuBr-AbsS}
\end{figure}

The absolute value $\vert\hat{\sigma}(\omega)\vert$ of the conductivity of \CuBr\ for different temperatures is plotted in Fig.~\ref{fig:CuBr-AbsS}(a,b) in a double logarithmic fashion. Corresponding to the non-metallic resistivity [Fig.~\ref{fig:resistivity}(a)], the conductivity increases with frequency at $T>100$~K. For lower temperatures, the curves approach a power-law in both directions: $\vert\sigma(\omega)\vert \propto \omega^{-1}$.  The phase angle of conductivity $\phi(\omega)$ shows a change in sign and eventually saturates at around $80^{\circ}$ when low temperatures are reached.
It is interesting to note that below 1000~\cm\ the phase angle changes only slightly with frequency,
dominant variations occur in the range 1000 - 3000~\cm, but they significantly shift $\phi(\omega)$
at low frequencies.

\begin{figure}[h!]
    \centering
        \includegraphics[width=0.8\linewidth]{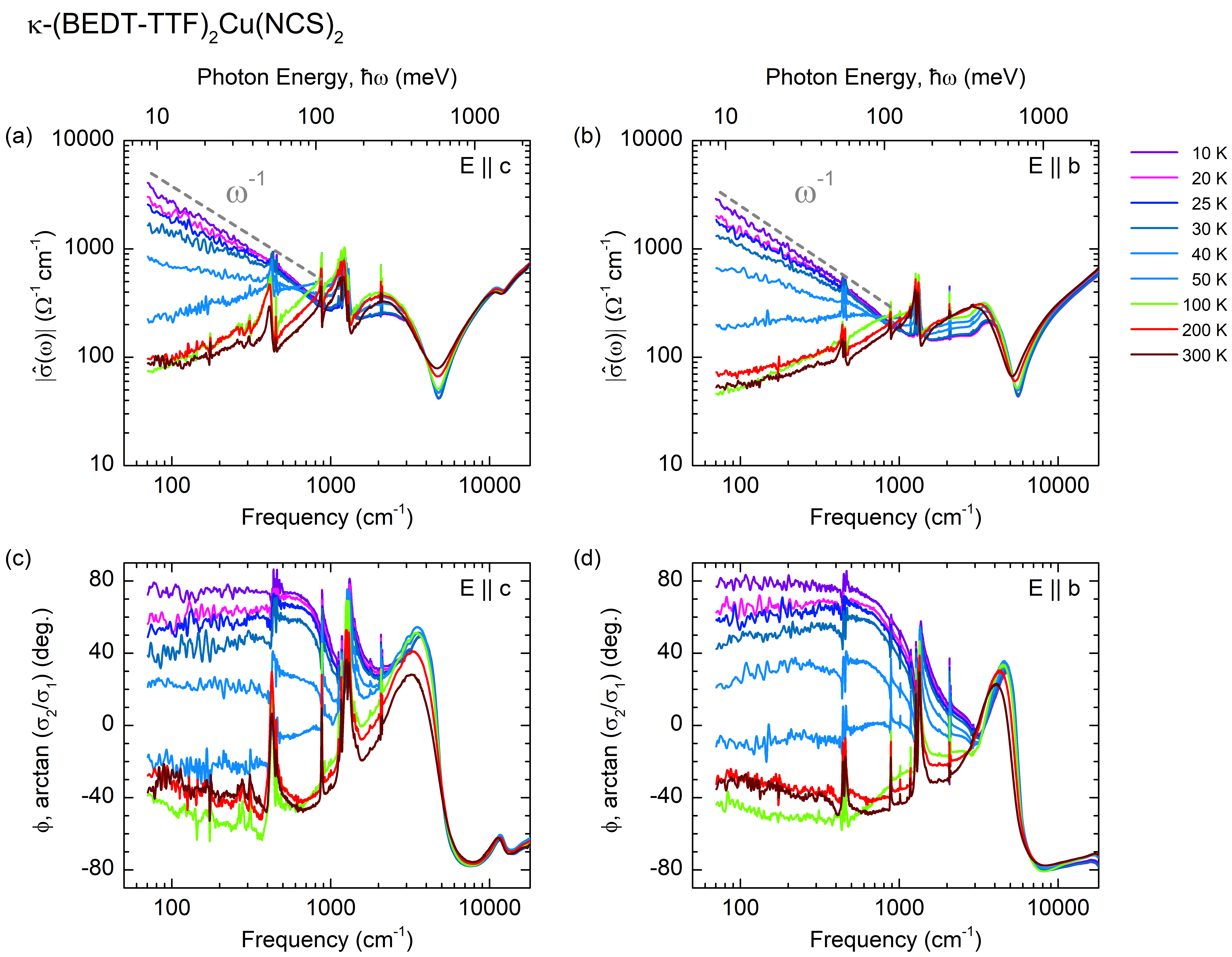}
        \caption{\textbf{Absolute value of optical conductivity, $\vert\hat{\sigma}(\omega)\vert$, and phase angle, $\phi(\omega)$, of \kCuNCS:} (a,b) Temperature-dependent absolute value of optical conductivity, for the two in-plane polarization axes ($E \parallel c$ and $E \parallel b$) for \CuNCS, illustrates that at low temperatures and low frequencies, $\vert\hat{\sigma}\vert \propto \omega^{-1}$; and (c,d) the phase angle, $\phi(\omega)$, values are close to 80$^\circ$.}
    \label{fig:CuNCS-AbsS}
\end{figure}

The corresponding plots of the optical properties of \CuNCS\ are presented in Fig.~\ref{fig:CuNCS-Ref} and Fig.~\ref{fig:CuNCS-AbsS}; the observations are similar to those obtained for \CuBr.
The findings for the temperature-dependent $R(\omega)$ and $\sigma_1(\omega)$ for \CuNCS, are also consistent with previous reports Kornelsen {\it et al.} \cite{Kornelsen91}.
Note, the different labeling of the crystallographic axes for \CuNCS\ compared to \CuBr\ due to the distinct crystal structure, as discussed in Sec.~\ref{sec:structureFermiliquidsystems}.

For further investigation and comparison to Hg-incommensurate systems, we consider the optical response at low temperatures (below 50 K) for both \CuBr{} and \CuNCS, where these compounds do attain the metallic and eventually Fermi-liquid behavior \cite{Kanoda06, Dumm09}.

\clearpage

\begin{figure}[h!]
    \centering
        \includegraphics[width=0.75\linewidth]{SI-CuNCS-Ref-S1-S2.png}
        \caption{\textbf{Optical response of \kCuNCS:} (a,b) Experimentally measured optical reflectivity, $R(\omega)$, of \CuNCS, in the temperatures range between 10 and 300 K for two in-plane polarization axes, $E \parallel c$ and $E \parallel b$; calculated (c,d) $\sigma_1(\omega)$, (e,f) $\sigma_2(\omega)$, and (g,h) $\varepsilon_1(\omega)$. The low-energy extrapolation in reflectivity (below 70 \cm) in (a,b) is indicated by dashed lines. Below 50 K, \CuNCS{} exhibits drastic changes in the optical response, below 3000 \cm, as is evident from sharp increase in the values of $\sigma_1$ and $\sigma_2$, and observation of zero-crossing around 2000-3000 \cm{} in $\varepsilon_1$, marking the metallic response below 50 K for \CuNCS.}
    \label{fig:CuNCS-Ref}
\end{figure}

\clearpage

\clearpage

\subsubsection{Strange metals}
\label{sec:opticsstrangemetals}
In Fig.~\ref{fig:HgBr-Ref} the optical properties of the incommensurate \HgBr\ are plotted at various temperatures. The corresponding plots for \HgCl\ are displayed in Fig.~\ref{fig:HgCl-Ref}.
Already from the reflectivity data, clear differences are seen in contrast to the Fermi-liquid metals \CuBr\ and \CuNCS, shown Fig.~\ref{fig:CuBr-Ref} and \ref{fig:CuNCS-Ref}. The $R(\omega)$ of the canonical metals exhibits a pronounced plasma edge, while the strange metals, \HgBr\ and \HgCl, show a much more gradual decrease, which approaches a linear drop in frequency as the temperature is lowered. This becomes particularly obvious in the low-frequency range, where the itinerant electrons dominate the optical response. Molecular vibrations are prominent at frequencies above 1200~\cm.

The molecular vibrations around 1200~\cm\ and the absorption bands in $\sigma_1(\omega)$ centred at 3000 \cm\ are shown in Fig.~\ref{fig:HgBr-Ref}(c,d) for \HgBr\ and for \HgCl\ in Fig. \ref{fig:HgCl-Ref}(c,d).
They are similar to interband excitations present in the mid-infrared region of \CuBr, and \CuNCS,
as seen in Figs.~\ref{fig:CuBr-Ref}(c,d) and \ref{fig:CuNCS-Ref}(c,d).
The Fermi-liquid metals possess a clear shift of spectral weight from the mid-infrared range towards lower frequency as the temperature is reduced, eventually building up the Drude-like response; this is thoroughly discussed in Ref.~\cite{Faltermeier07,Dumm09}. In the present case of the strange metals \HgBr\ and \HgCl\ we do not find a clear spectral-weight transfer towards low frequencies upon cooling.
The observed variation in the mid-infrared features evolve independently from the low-frequency response. For that reason, here we confine our discussion to low frequencies.
The high-temperature optical spectra (above $T\approx 150$~K) is different compared to other $\kappa$-BEDT-TTF systems, and potentially much more complex than isostructural compounds, like \CuBr{} and \CuNCS. We will not discuss this behavior in our current work.

We want to note that for none of the four compounds, $\sigma_2(\omega)$ spectra exhibit a maximum
with a drop towards lower energies. This implies that our experimental frequency window
remains well in the relaxation regime \cite{Dressel02Book}.

\begin{figure}[!ht]
    \centering
        \includegraphics[width=0.73\linewidth]{SI-HgBr-Ref-S1-S2.png}
        \caption{\textbf{Optical response of \kHgBrd:} (a,b) Experimentally measured optical reflectivity, $R(\omega)$, of \HgBr, in the temperatures range between 10 and 300 K for the two in-plane polarization axes, $E \parallel b$ and $E \parallel c$, (c,d) optical conductivity $\sigma_1(\omega)$, (e,f) $\sigma_2(\omega)$, and (g,h) $\varepsilon_1(\omega)$ calculated via the Kramers-Kronig relations. The low-energy extrapolation in reflectivity (below 35 \cm) in (a,b) is indicated by dashed lines. \HgBr{} demonstrates metallic response in conductivity down to 10~K, with small, but finite $\sigma_1$, $\sigma_2$ values even at 300 K, and zero-crossing observed in $\varepsilon_1$.  The low frequency response ($<$ 2000 \cm) evolves systematically, and independent of any pronounced spectral weight transfer from the absorption bands at higher frequencies.}
    \label{fig:HgBr-Ref}
\end{figure}
\clearpage

\begin{figure}[!ht]
    \centering
        \includegraphics[width=0.8\linewidth]{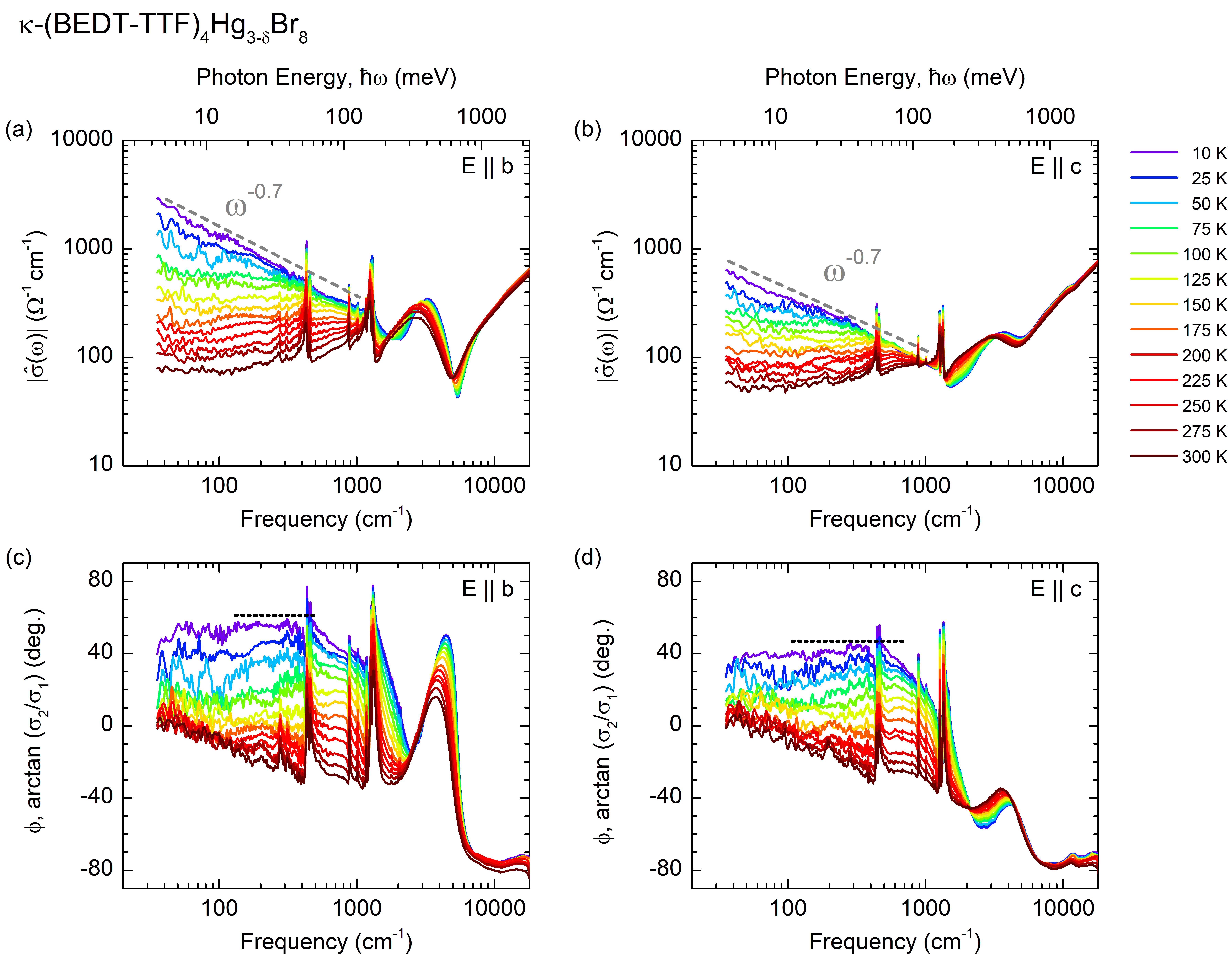}
        \caption{\textbf{Absolute value of optical conductivity, $\vert\hat{\sigma}(\omega)\vert$, and phase angle, $\phi(\omega)$, of \kHgBrd:} (a,b) Temperature-dependent absolute value of optical conductivity, for the two in-plane polarization axes ($E \parallel b$ and $E \parallel c$) of the strange metal \HgBr, illustrates that at low temperatures and low frequencies $\vert\hat{\sigma}(\omega)\vert \propto \omega^{-0.7}$; and (c,d) the phase angle $\phi(\omega)$, values are close to 60$^\circ$.}
    \label{fig:HgBr-AbsS}
\end{figure}

\clearpage
\begin{figure}[!ht]
    \centering
        \includegraphics[width=0.75\linewidth]{SI-HgCl-Ref-S1-S2.png}
        \caption{\textbf{Optical response of \kHgCld:} (a,b) Experimentally measured optical reflectivity, $R(\omega)$, of \HgCl, in the temperatures range between 10 and 300 K for two in-plane polarization axes, $E \parallel b$ and $E \parallel c$; calculated (c,d) $\sigma_1(\omega)$, (e,f) $\sigma_2(\omega)$, and (g,h) $\varepsilon_1(\omega)$. The low-energy extrapolation in reflectivity (below 40 \cm) in (a,b) is indicated by dashed lines. The optical response of \HgCl{} is very similar to that of \HgBr, and demonstrates but finite values of $\sigma_1$, $\sigma_2$ values upto 300 K, and zero-crossing in $\varepsilon_1$.  The low frequency response ($<$ 2000 \cm) evolves systematically, and independent of any pronounced spectral weight transfer from the absorption bands at higher frequencies. \HgCl{} records additional changes at low frequencies below $T\approx 20$~K, which are consistent with changes slope in resistivity.}
    \label{fig:HgCl-Ref}
\end{figure}
\clearpage

\begin{figure}[h!]
    \centering
        \includegraphics[width=0.8\linewidth]{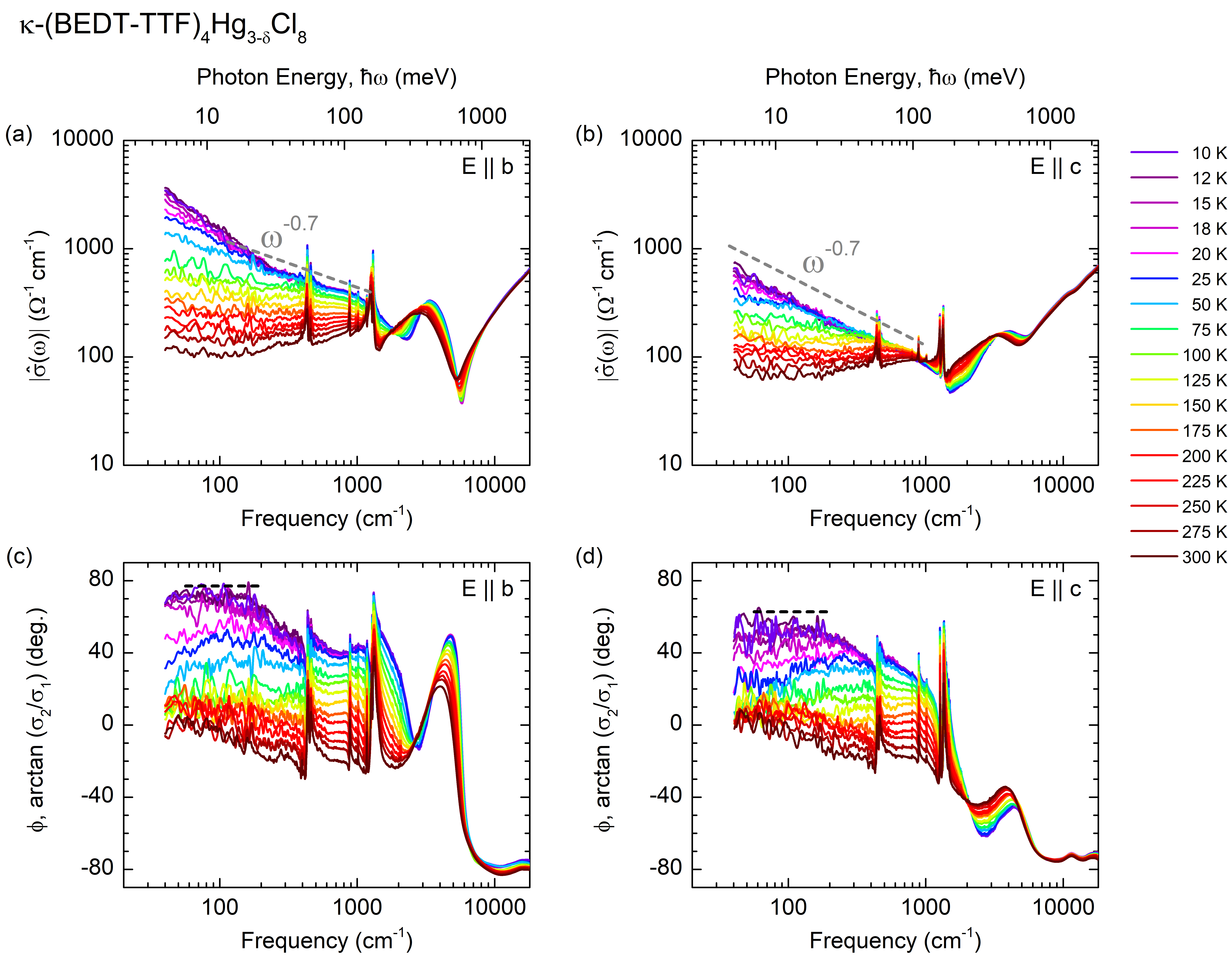}
        \caption{\textbf{Absolute value of optical conductivity, $\vert\hat{\sigma}(\omega)\vert$, and phase angle, $\phi(\omega)$, of \kHgCld:} (a,b) Temperature-dependent absolute value of optical conductivity, for the two in-plane polarization axes ($E \parallel b$ and $E \parallel c$)  for \HgCl, illustrates that at low temperatures and low frequencies, $\vert\hat{\sigma}(\omega)\vert \propto \omega^{-0.7}$; and (c,d) the phase angle, $\phi(\omega)$, values are close to 60$^\circ$.}
    \label{fig:HgCl-AbsS}
\end{figure}

In Figs.~\ref{fig:CuBr-AbsS} and \ref{fig:CuNCS-AbsS} we have seen that $\vert\hat{\sigma}(\omega)\vert$ {\it vs.} $\omega$ for the Fermi liquids \CuBr\ and \CuNCS\ attains a $\omega^{-1}$ response, with the values of the phase angle, $\phi(\omega)$, around 80$^\circ$ (below 1000~\cm).
Figs.~\ref{fig:HgBr-AbsS} and \ref{fig:HgCl-AbsS} display the data for the incommensurate compounds \HgBr{} and \HgCl\ in a similar fashion.
For the strange metals the frequency dependent $\vert\hat{\sigma}(\omega)\vert$ response obtains a $\omega^{-0.7}$ dependence with the $\phi(\omega)$ of approximately 60$^\circ$ (below 1000~\cm),  which are typical observations from cuprate superconductors in their strange metals \cite{Marel03, Michon23}.
This experimental facts can be phenomenologically described as the stretched Drude response,
due to change of $\omega^{-1}$ exponential term to $\approx \omega^{-0.7}$.
The itinerant carriers experience an extended response that can converge to theoretically calculated conditions for strange metal and the accompanying Planckian dissipation in the frequency dependence \cite{Michon23,Fratini26}.

\clearpage

\subsection{Comparison of electrical resistivity obtained by transport measurement and by optical investigations: $\rho_{\mathrm dc}$ {\it vs.} $\rho_{\omega_c}$}

\begin{figure}[h!]
    \centering
        \includegraphics[width=0.6\linewidth]{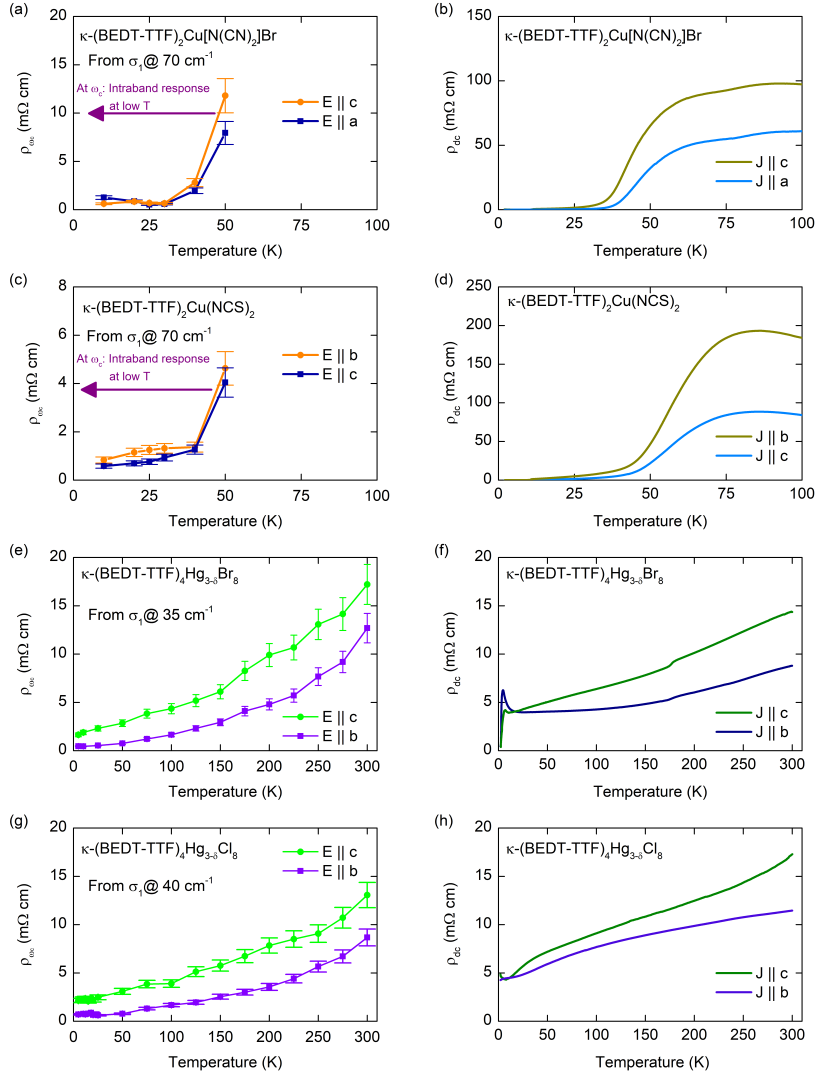}
        \caption{\textbf{Comparison of dc resistivity and resistivity from the optical measurements: } $\rho_{\omega_c}$ is calculated at the lower frequency cut-offs for the four compounds under investigation. $\rho_{\mathrm dc}$ is taken from four-probe transport measurements:
        (a,b) \CuBr, (c,d) \CuNCS, (e,f) \HgBr, and (g,h) \HgCl.
        For \CuBr\ and \CuNCS\ only relevant temperatures below 50 K are shown.}
    \label{fig:Res-S1}

\end{figure}

Here we validate that our frequency-dependent measurements capture the electronic ground state
and that the relaxation of itinerant carriers is not affected by any other factors.
As already mentioned above in Sec.~\ref{sec:opticsstrangemetals}, we do not have any indication
of a peak in $\sigma_2(\omega)$ at low energies, where the Drude model (Eq. \ref{eq:Drude-Sommerfeld}) expects a maximum at $\omega\tau = 1$.
In the following, we compare the dc resistivity, $\rho_{\mathrm dc}$, obtained from our transport studies
with the optical resistivity in the low-frequency limit. To that end, we calculate
$\rho_{\omega_c}$ from the optical conductivity  $\sigma_1(\omega)$ at the lowest measured frequency $\omega_c$, which differs slightly due to sample size and data quality.
Fig.~\ref{fig:Res-S1} presents the respective values as a function of temperature;
the cutoff frequency is indicated for each sample.

For the Fermi-liquid compounds, \CuBr\ and \CuNCS, as shown in  Fig.~\ref{fig:Res-S1}(a-d),
the comparison of $\rho_{\omega_c}$ and $\rho_{\mathrm dc}$ is limited to the Fermi-liquid behavior below 50~K, where we find qualitative agreement.

For strange metals \HgBr\ and \HgCl, presented in Fig.~\ref{fig:Res-S1}(e-h),
$\rho_{\omega_c}$ exhibits comparable absolute values and the same anisotropy as obtained from dc measurements. The metallic state is characterized by linear increase of resistivity
over a large temperature range.

\clearpage

\subsection{Extended Drude analysis}

Going beyond the conventional Drude model, the extended Drude model extracts a frequency-dependent scattering rate $\Gamma(\omega)$ and effective mass $m^{*}$, from Eq. (\ref{eq:Drude-Sommerfeld}), without any underlying assumptions \cite{Dressel02Book}:
\begin{equation}
\hat{\Gamma}(\omega) = \Gamma_1(\omega) + {\mathrm i}\Gamma_2(\omega) \quad,
\label{eq:gamma-complex}
\end{equation}
\begin{equation}
\Gamma(\omega) = \frac{\omega_{\mathrm{p}}^{2}}{4\pi} \frac{\sigma_{1}(\omega)}{|\hat{\sigma}(\omega)|^{2}}  \quad  {\mathrm{and}} \quad \frac{m^*}{m_b} = \frac{\omega_{\mathrm{p}}^{2}}{4\pi}  \frac{\sigma_{2}(\omega)/\omega}{|\hat{\sigma}(\omega)|^{2}} \quad,
\label{eq:gamma-m}
\end{equation}
where the plasma frequency is given as $\omega_p=\sqrt{4\pi N e^2/m}$, with the carrier density $N$, the elementary charge $e$ and charge carrier mass $m$. $\hat{\Gamma}(\omega)$ is the frequency-dependent complex scattering rate, containing the real part $\Gamma_1(\omega)$ [sometimes simply denoted as $\Gamma(\omega)$] and the imaginary part $\Gamma_2(\omega)$, which is related to the effective mass by: $m^*(\omega)/m_b = 1-\Gamma_2(\omega)/\omega$, with band mass $m_b$. The plasma frequency $\omega_{\mathrm{p}}$ was obtained from the zero-crossing of the real part of the dielectric constant:
\begin{equation}
\omega_{\mathrm{p}} = \omega_{\rm screened} \cdot \sqrt{\varepsilon_{\infty}} \quad ,
\label{eq:wp}
\end{equation}
where $\omega_{\rm screened}$ is the plasma frequency screened due to interband transitions, and $\varepsilon_{\infty}$ is obtained from the high-frequency limit (from fit) of the dielectric constant at a cut-off frequency $\omega_{\rm cut-off}$ = 50\,000~\cm.

Landau's Fermi-liquid theory predicts a renormalized energy-dependent scattering rate:
\begin{equation}
\frac{1}{\tau^{*}(T,\omega)} = \Gamma (T,\omega) \varpropto [ (2 \pi k_{B}T)^{2} + (\hbar\omega)^{2}]  \quad,
\label{eq:FermiLiq-gamma}
\end{equation}
where $k_{B}$ is the Boltzmann constant, $T$ is the temperature, and $\hbar$ is the reduced Planck constant. Eq.~(\ref{eq:FermiLiq-rhodc})
is commonly utilized to describes the $T^2$ dependence of the dc resistivity 
in the Fermi liquid state:
\begin{equation}
\rho_{\mathrm{dc}} =  \left( \frac{N e^2}{m^{*}} \tau^{*} \right)^{-1} \approx \frac{m^{*}}{N e^2} 2 \pi k_{B}^{2} T^{2} \quad.
\label{eq:FermiLiq-rhodc}
\end{equation}

The frequency dependence of the low-energy electron dynamics has been discussed by Berthold {\it et al.}~\cite{Berthod13}.

The linear-in-temperature increase of the dc resistivity at low temperature is accounted for by
replacing the scattering rate in the conventional Drude model (\ref{eq:Drude-Sommerfeld}) by the Planckian relation $k_{B}T/\hbar$, resulting in a ``Planckian metal'' \cite{Patel19}. Inserting
\begin{equation}
\frac{1}{\tau} = \alpha \frac{k_{B}T}{\hslash}
\label{eq:Planckian}
\end{equation}
into the resistivity $\rho_{\mathrm{dc}} = m/(N e^2 \tau)$, we arrive at
\begin{equation}
\rho_{\mathrm{dc}} = \alpha \frac{m}{N e^2} \frac{k_{B}T}{\hslash} \quad ,
\label{eq:Planckian}
\end{equation}
where $\alpha$ is a sample-dependent prefactor, generally on the order of unity.

In this section, we present the complete set of data for all four compounds under investigations, for both polarizations and for different temperatures as indicated.
We confine this analysis for low temperatures, to investigate the coherent itinerant carrier behavior.

\begin{figure}[h!]
    \centering
        \includegraphics[width=0.9\linewidth]{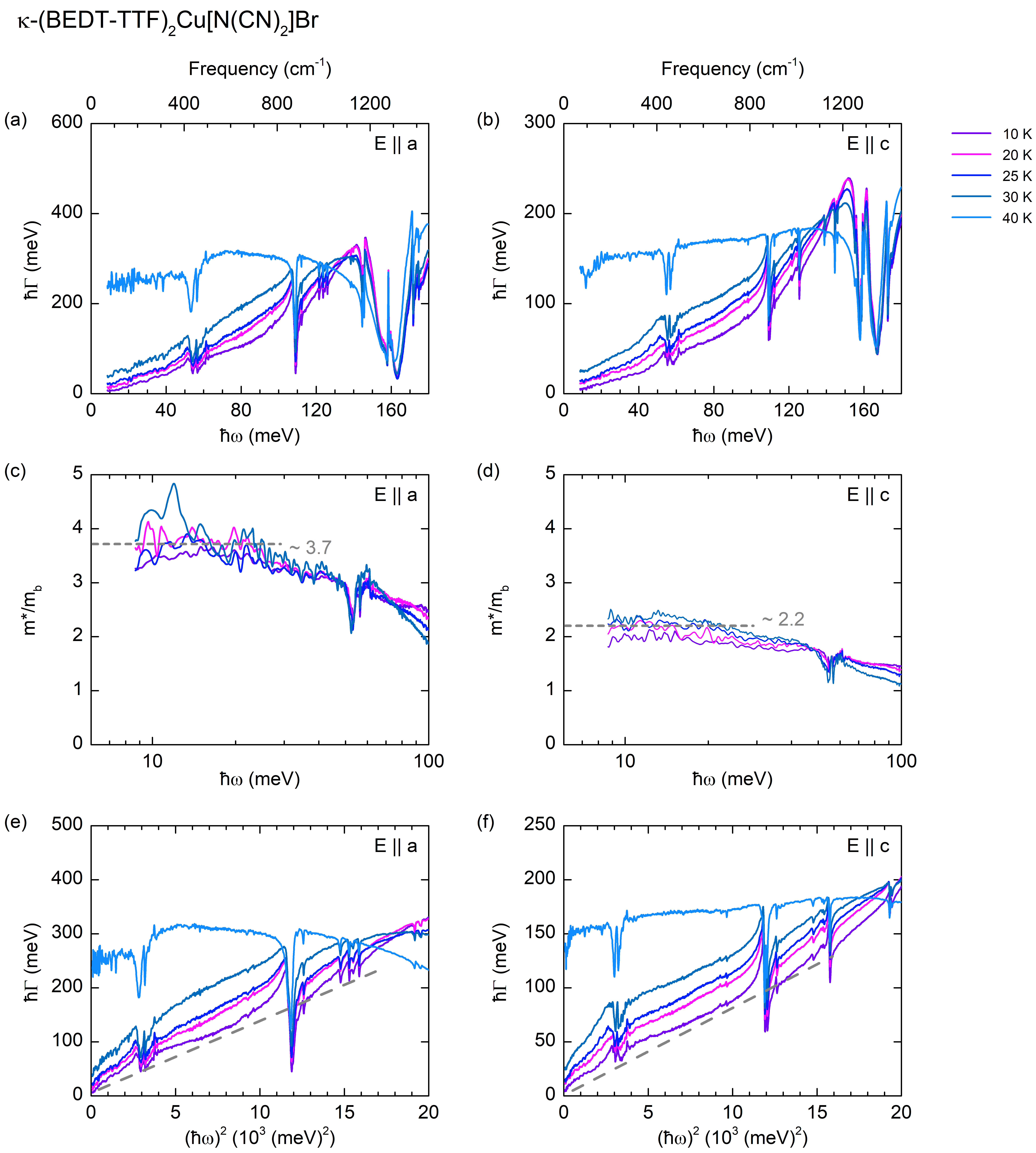}
        \caption{\textbf{Frequency-dependent scattering rate and Fermi-liquid behavior for \kCuBr:}  (a,b) Frequency-dependent scattering rate $\Gamma(\omega,T)$ {\it vs.} photon energy $\hslash\omega$ for the two in-plane axes for \CuBr\ at different temperatures. Since we focus on the itinerant electrons, we confine ourselves to low energies. The negative slope at high temperatures, above 100 K, is due to formation of an absorption band, instead of the coherent intraband response, observed only at low temperatures. (c,d) The effective mass yields values for 3.7 and 2.2 for the two different polarization axes, when extrapolated with constant values towards low energies. (e,f) $\Gamma(\omega,T)$ \textit{vs.} $(\hslash\omega)^2$ shows linear dependence at low temperatures, confirming the Fermi-liquid behavior in \CuBr. The sharp asymmetric features are the molecular vibrations, which are not removed from the optical conductivity spectra, to avoid inducing artificial features in the experimental results. The dashed lines in panels (c,d) are guides for eye to show the linear dependence at low energies.}
    \label{fig:CuBr-ExtDrude}
\end{figure}

\begin{figure}[h!]
    \centering
        \includegraphics[width=0.9\linewidth]{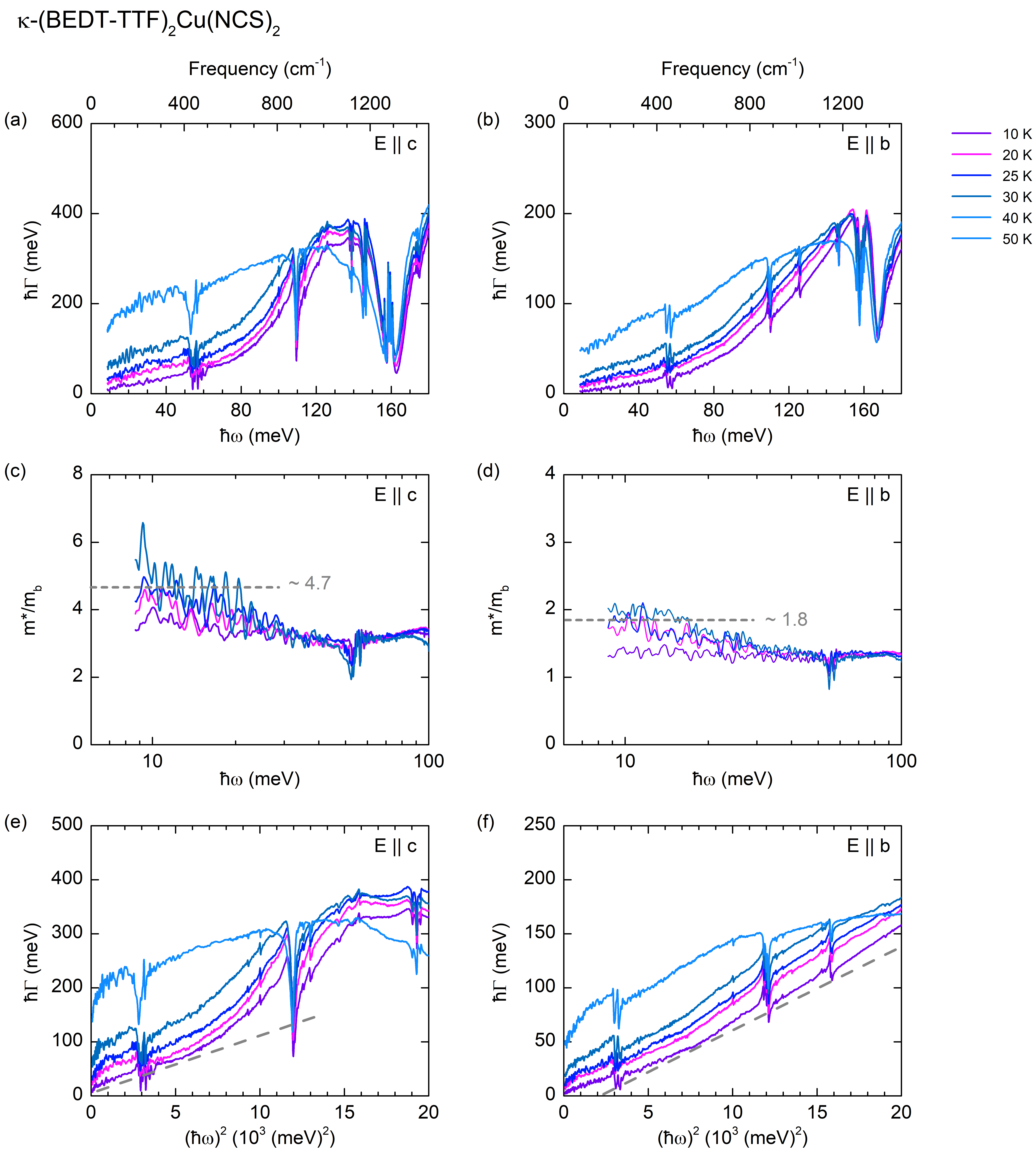}
        \caption{\textbf{Frequency-dependent scattering rate and Fermi-liquid behavior for \kCuNCS:}  (a,b) Frequency-dependent scattering rate, $\Gamma(\omega,T)$ {\it vs.} photon energy, $\hslash\omega$, for the two in-plane axes for \CuNCS{} for different temperatures, at low frequencies. The negative slope at high temperatures, above 50 K, is due to formation of an absorption band, instead of the coherent intraband response, observed only at low temperatures. (c,d) The effective mass calculated from the extended Drude analysis gives values for 4.7 and 1.8 for the two different polarization axes, when extrapolated with constant values towards low energies. (e,f) $\Gamma(\omega,T)$ \textit{vs.} $(\hslash\omega)^2$ shows linear dependence at low temperatures, confirming the Fermi-liquid behavior in \CuNCS. The sharp asymmetric features are the molecular vibrations, which are not removed from the optical conductivity spectra, to avoid inducing artificial features in the experimental results. The dashed lines in panels (e,f) are guides for eye to show the linear dependence at low energies.}
    \label{fig:CuNCS-ExtDrude}
\end{figure}

In the case of the Fermi-liquid compounds, \CuBr{} and \CuNCS, as presented in Figs. \ref{fig:CuBr-ExtDrude} and \ref{fig:CuNCS-ExtDrude}, $\Gamma \propto \omega^2$, in accordance with the Fermi-liquid theory [Eq.~(\ref{eq:FermiLiq-gamma})].

\begin{figure}[h!]
    \centering
        \includegraphics[width=0.85\linewidth]{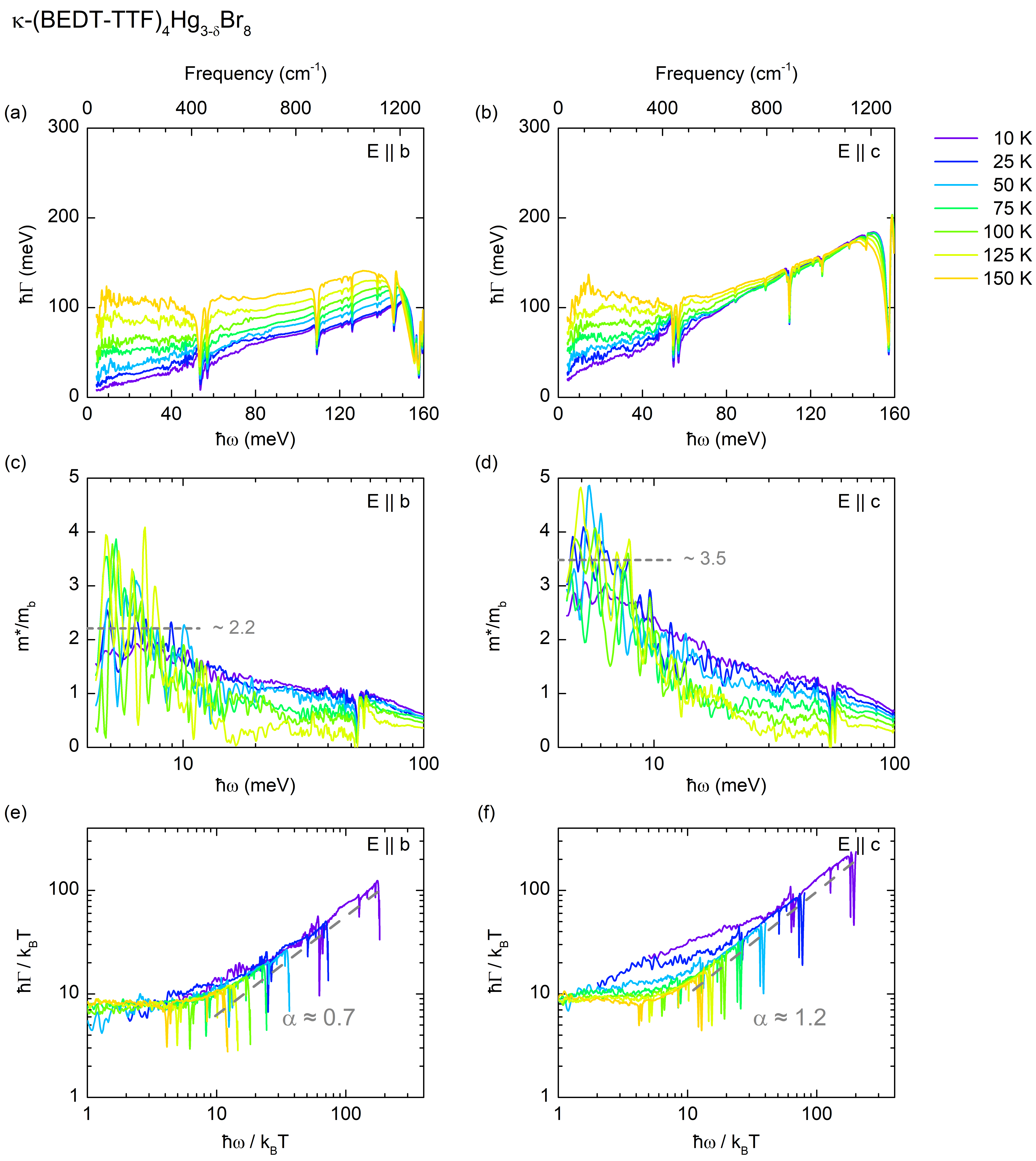}
        \caption{\textbf{Frequency-dependent scattering rate and Planckian scaling for \kHgBrd:}  (a,b) Frequency-dependent scattering rate, $\Gamma_{1}(\omega,T)$ {\it vs.} photon energy, $\hslash\omega$ for the two in-plane axes for \HgBr{} for different temperatures, at low frequencies. (c,d) The effective mass calculated from the generalized Drude analysis, which gives values for 2.2 and 3.5 for the two different polarization axes, when extrapolated with constant values towards low energies. (e,f) The low energy relevant region from the intraband response, illustrating the linear response observed with the Planckian dissipation scaling, where the spectra at different temperatures fall onto a single line for \HgBr. The high energy values are removed from display for clarity. The dashed lines in panels (e,f) are the linear fits, corresponding to the $\alpha$ parameter observed in the Planckian scaling (presented with an offset for clarity).}
    \label{fig:HgBr-ExtDrude}
\end{figure}

\begin{figure}[h!]
    \centering
        \includegraphics[width=0.85\linewidth]{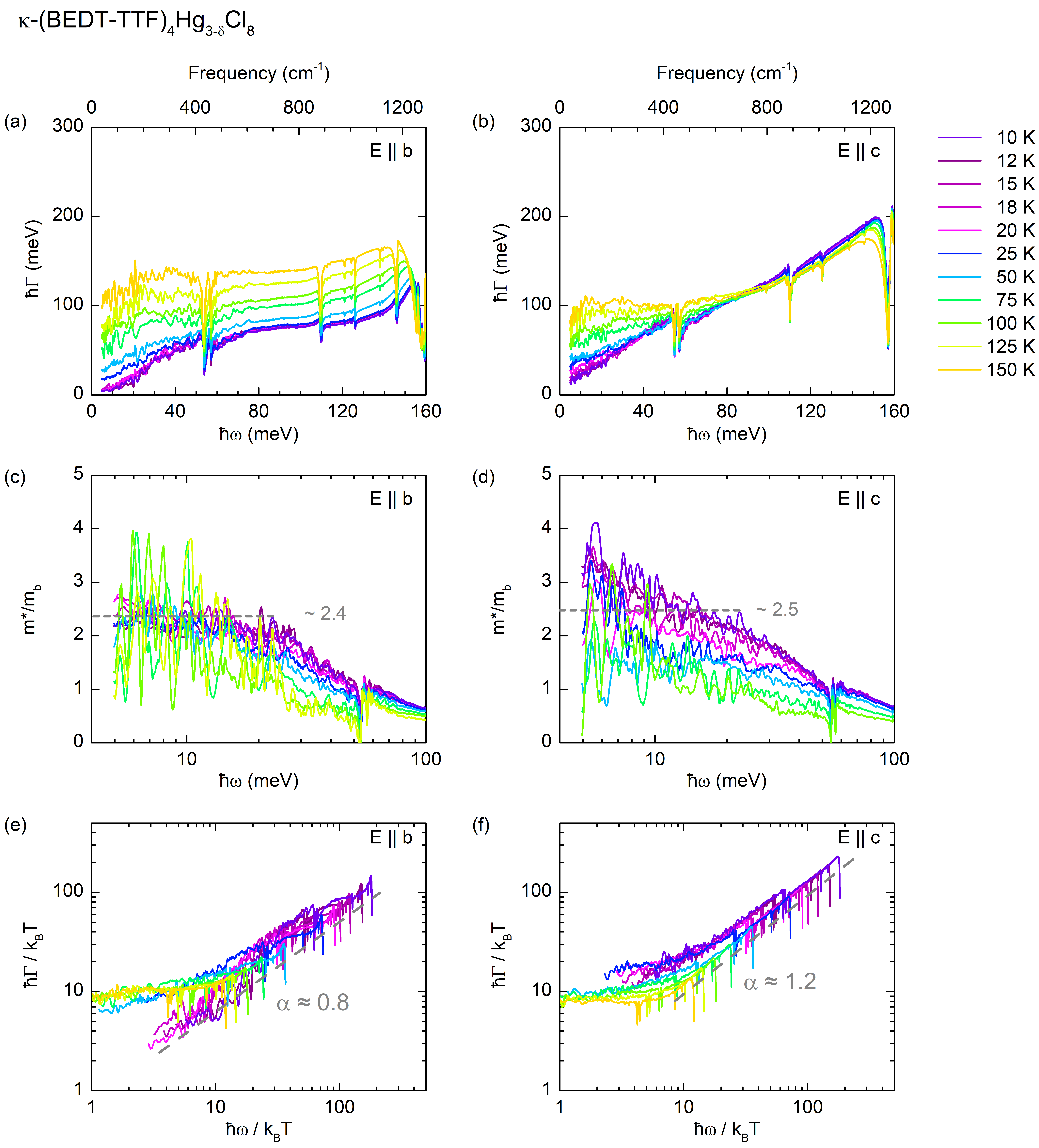}
        \caption{\textbf{Frequency-dependent scattering rate and Planckian scaling for \kHgCld:}  (a,b) Frequency-dependent scattering rate, $\Gamma_{1}(\omega,T)$ {\it vs.} photon energy, $\hslash\omega$ for the two in-plane axes for \HgCl{} for different temperatures, at low energies. (c,d) The frequency-dependent effective mass gives values for 2.4 and 2.5 for the two different polarization axes, when extrapolated with constant values towards low energies. (e,f) The low energy relevant region from the intraband response, illustrating the linear response observed with the Planckian dissipation scaling, where the spectra at different temperatures fall onto a single line, similar to \HgBr. The high energy values are removed from display for clarity. The dashed lines in panels (e,f) are the linear fits, corresponding to the $\alpha$ parameter observed in the Planckian scaling (presented with an offset for clarity).}
    \label{fig:HgCl-ExtDrude}
\end{figure}

For the strange-metal compounds, \HgBr{} and \HgCl, as illustrated in Figs. \ref{fig:HgBr-ExtDrude} and \ref{fig:HgCl-ExtDrude}, $\Gamma \propto \omega$ and can be scaled to the Planckian dissipation relations, generally employed for transition metal oxide strange metals \cite{Michon23}.

\clearpage

\clearpage

\clearpage

\newpage
\bibliography{ref}